# Thermodynamic Stability of Molybdenum Oxycarbides Formed from Orthorhombic $Mo_2C$ in Oxygen-rich Environments

S.R.J. Likith,[a,†] C. A. Farberow,[b,*] S. Manna,[a,†] A. Abdulslam,[a] V. Stevanović,[c,d] D. A. Ruddy,[b] J. A. Schaidle,[b] D. J. Robichaud,[b,*] and C.V. Ciobanu[a,*]

[a] *Department of Mechanical Engineering, Colorado School of Mines, Golden, CO 80401, USA*

[b] *National Bioenergy Center, National Renewable Energy Laboratory, Golden, CO 80401, USA*

[c] *National Renewable Energy Laboratory, Golden, CO 80401, USA*

[d] *Department of Metallurgical and Materials Engineering, Colorado School of Mines, Golden, CO, 80401, USA*

**Abstract:** Molybdenum carbide ($Mo_2C$) nanoparticles and thin films are particularly suitable catalysts for catalytic fast pyrolysis (CFP) as they are effective for deoxygenation and can catalyze certain reactions that typically occur on noble metals. Oxygen deposited during deoxygenation reactions may alter the carbide structure leading to the formation of oxycarbides, which can determine changes in catalytic activity or selectivity. Despite emerging spectroscopic evidence of bulk oxycarbides, so far there have been no reports of their precise atomic structure or their relative stability with respect to orthorhombic $Mo_2C$. This knowledge is essential for assessing the catalytic properties of molybdenum (oxy)carbides for CFP. In this article, we use density functional theory (DFT) calculations to (a) describe the thermodynamic stability of surface and subsurface configurations of oxygen and carbon atoms for a commonly studied Mo-terminated surface of orthorhombic $Mo_2C$, and (b) determine atomic structures for oxycarbides with a Mo:C ratio of 2:1. The surface calculations suggest that oxygen atoms are not stable under the top Mo layer of the $Mo_2C(100)$ surface. Coupling DFT calculations with a polymorph sampling method, we determine $(Mo_2C)_xO_y$ oxycarbide structures for a wide range of oxygen compositions. Oxycarbides with lower oxygen content ($y/x \leq 2$) adopt layered structures reminiscent of the parent carbide phase, with flat Mo layers separated by layers of oxygen and carbon; for higher oxygen content, our results suggest the formation of amorphous phases, as the atomic layers lose their planarity with increasing oxygen content. We characterize the oxidation states of Mo in the oxycarbide structures determined computationally, and simulate their X-ray diffraction (XRD) patterns in order to facilitate comparisons with experiments. Our study may provide a platform for large-scale investigations of the catalytic properties of oxycarbides and their surfaces, and for tailoring the catalytic properties for different desired reactions.

† These authors made equal contributions to the results in this work

* To whom correspondence may be addressed: carrie.farberow@nrel.gov (C. A. Farberow), david.robichaud@nrel.gov (D. J. Robichaud), and cciobanu@mines.edu (C.V. Ciobanu)

[†] These authors made equal contributions to the results in this work

[*] To whom correspondence may be addressed: carrie.farberow@nrel.gov (C. A. Farberow), david.robichaud@nrel.gov (D. J. Robichaud), and cciobanu@mines.edu (C.V. Ciobanu)

## 1. Introduction

Catalytic fast pyrolysis (CFP) is a viable thermochemical conversion route for producing fuels from biomass.[1-3] One approach to obtaining stable bio-oils is through ex situ CFP, which involves sending pyrolysis vapors to a downstream reactor where oxygen functionalities are removed (deoxygenation) through the use of multifunctional catalysts.[4-5] The development of more active selective, inexpensive, and recyclable catalysts for ex situ CFP can improve the yield and properties of upgraded bio-oils. In this context, transition metal carbides, in particular molybdenum carbide, have emerged as promising materials that can convert a wide range of oxygenates such as acetic acid,[6-9] guaiacol,[10-12] and phenol,[13] often in the presence of water.[13-15] Scientific and technologic interest in synthesizing molybdenum carbides has been fueled by the discovery that they are active for many of the reactions that usually rely on noble metal catalysts,[16-18] making these carbides appealing catalytic materials with certain economic advantages. Molybdenum carbide catalysts are active for hydrogenation,[19-21] Fischer-Tropsch reactions,[22-23] and methane reforming,[24] among others.[6-13] The catalytic performance of molybdenum carbides depends on the crystal structure, as well as on the exact orientation and termination of the exposed surfaces.[1,25-28] Various procedures can be used to synthesize molybdenum carbide, including reactions with metal compounds,[29-32] pyrolysis of metal precursors,[33] and carburization of molybdenum oxide.[34-37]

In one of the most widely used synthesis methods, carburization of molybdenum oxide, the choice of carbon source can lead to a wide range of structures and morphologies.[38-40] In addition, reaction conditions, such as temperature, pressure, and reaction time, affect the degree of carburization and may result in a final product containing residual oxygen in the form of Mo oxycarbides with varying oxygen and carbon content. Even when carbide formation is complete, the catalytic operation of molybdenum carbide thin films or nanoparticles in an environment rich in oxygen-containing molecules (e.g., $O_2$, $H_2O$, CO, $CO_2$, organic oxygenates) under specific reaction conditions can lead to the formation of oxycarbides, which have been hypothesized to exist at or near the surface.[41-43] Surface oxycarbides also form when the carburized material is passivated in oxygen environment.[44-45] Given the propensity for oxycarbide formation during catalyst synthesis or operation in environments with oxygen-containing reactants and products, the bulk and surface structure may differ significantly from those of the initial carbide, and in turn trigger changes in the catalytic behavior as well. Clear spectroscopic evidence of bulk oxycarbide phases has recently been reported by Óvári et al.,[46] who studied the interaction of molecular oxygen with the (0001) surface of $Mo_2C$, and found that above 800K there was a change in the oxidation states of the Mo atoms. Such change is indicative of oxygen migration into the bulk at least down to the interrogation depth (~6 nm) of X-ray photoelectron spectroscopy (XPS). This oxygen migration, accompanied by reported variations in the oxidation state[46-47] of Mo, indicates structural changes. Thus, these experiments[46] provide evidence of bulk oxycarbide formation during high temperature exposure of $Mo_2C$ to molecular oxygen. Despite indications of bulk Mo oxycarbide phases and evidence that oxygen modification of Mo carbides changes their catalytic performance,[48-50] so far there have been only few (tentative) reports of their atomic structure[51-52] and stability. Computational modeling of reactions catalyzed by molybdenum carbide generally assumes oxygen-free bulk $Mo_2C$,[10,53-55] and sometimes the presence of oxygen atoms immediately above or below the surface layer is considered.[6,41,43] Knowledge of the bulk structure of oxycarbides is, however, essential for future assessments of catalytic properties of molybdenum (oxy)carbides for ex situ CFP, and can lead to advancements in catalyst multi-functionalization and regeneration.

In this article, we use first principles calculations based on density functional theory (DFT) to (i) describe the thermodynamic stability of surface and subsurface configurations of oxygen and carbon atoms for the Mo-terminated (100) surface of orthorhombic ($a < b < c$) $Mo_2C$, and (ii) find atomic structures for oxycarbides with a wide range of O compositions and a specific Mo:C ratio. The surface calculations model a common facet of $Mo_2C$ nanoparticles, while the bulk modeling is relevant for (polycrystalline) thin films, usually thicker than 100 nm.[56] For the surface calculations, we examined all the configurations with a set upper limit for the number of O and C atoms above and below the top Mo layer, and built a thermodynamic phase diagram showing the most stable surface phases for given thermodynamic conditions (chemical potentials). We found that any configuration with oxygen atoms below the surface is less stable than some counterpart with oxygen above the surface. This result suggests that either O atoms are unstable below the top Mo layer of $Mo_2C(100)$ regardless of reaction conditions, or the parent surface restructures significantly to accommodate oxygen atoms at high oxygen content. In order to further explore thermodynamic phase changes induced by oxygen, we sought low-energy oxycarbide structures using a polymorph sampling method. Depending on the oxygen content, we found that stable oxycarbides indeed exist, and become more stable than the parent phase with increasing chemical potential of oxygen. They adopt crystal structures markedly different than orthorhombic $Mo_2C$, and we characterized these structures computationally by determining their Mo oxidation states and simulating their X-ray diffraction (XRD) patterns. In practice, oxycabides may expose Mo-, C-, or O-terminated surfaces, which would exhibit different catalytic properties. Regardless of termination, the transformation from carbide to oxycarbide modifies the properties of all surface orientations that can be exposed by nanoparticles and thin films, but does not necessarily render them inactive. Due to the potential phase transformations suggested here, a $Mo_2C$ catalyst operating in an oxidizing environment could have different activity and selectivity while in operation as compared with its initial as-synthesized state. Given the diversity of oxycarbide bulk and surface structures, and the chemical reactions that can take place on the exposed surfaces, this work may open the door for future large scale studies of the catalytic properties of different oxycarbides and their surfaces, and tailoring of reactivity and selectivity for different desired reactions.

## 2. Computational Details

### 2.a. The $Mo_2C$ model surface and DFT slab calculations

The catalytic activity of any surface is determined by both its orientation and its termination. In the case of carbide nanoparticles, it is difficult to ascertain which facets are responsible for the observed activity. For the present work, we choose a stable, well-studied model surface, i.e., the Mo-terminated (100) surface of orthorhombic $Mo_2C$. Since there is a lack of consensus in naming this surface, we describe it here in the context of other reports in the literature. The Mo-terminated (100) surface used in this study is that of the orthorhombic phase of $Mo_2C$ (space group *Pbcn*),[57] with the lattice constants ordered as $a < b < c$. The 1×1 surface unit cell has dimensions $b \times c$, and the surface normal is oriented along the $a$ axis. Even though the crystal is orthorhombic, the Mo planes are approximately close-packed, which prompted the use of the 4-Miller index notation (0001) in some previous reports. [58-59] Because of differences in (re)ordering the values of the lattice constants and/or because the atoms in the Mo layers are hexagonally close packed with some distortion, this very same Mo-terminated surface has been denoted in the literature as $Mo_2C(001)$,[6,41-42] $\alpha$ –$Mo_2C(0001)$,[58] $\beta$ –$Mo_2C(001)$,[60-61] or $\beta$ –$Mo_2C(0001)$.[59]

Using DFT in the framework of the generalized gradient approximation (GGA), we have performed structural relaxations for slabs that expose the Mo-terminated $Mo_2C(100)$ surface, with varying numbers of O and C atoms placed above and below the top Mo layer. These DFT calculations were carried out using the Vienna Ab-Initio Simulation Package (VASP),[62] with the Perdew-Burke-Ernzerhof (PBE) exchange-correlation functional,[63] and projector-augmented wave (PAW) potentials.[64] We have obtained the lattice constants $a = 4.74884$ Å, $b = 5.25067$ Å, and $c = 6.07256$ Å, which are in good agreement with the experimental values[65] (4.72 Å, 5.20 Å, and 6.01 Å). The surfaces were modeled using 8-layer thick slabs (Fig. 1), with 1×1 periodically repeated unit cells, and vacuum spacing of 18 Å. The bottom two layers of the slabs were kept fixed, whereas atoms in and above the third layer were allowed to relax. Atomic relaxations with fixed periodic boundaries were carried out using dipole corrections and a convergence criterion for residual forces of 0.02 eV/Å. The plane-wave energy cutoff was set at 500 eV and the Brillouin zone was sampled using a Monkhorst-Pack 5×5×1 grid.

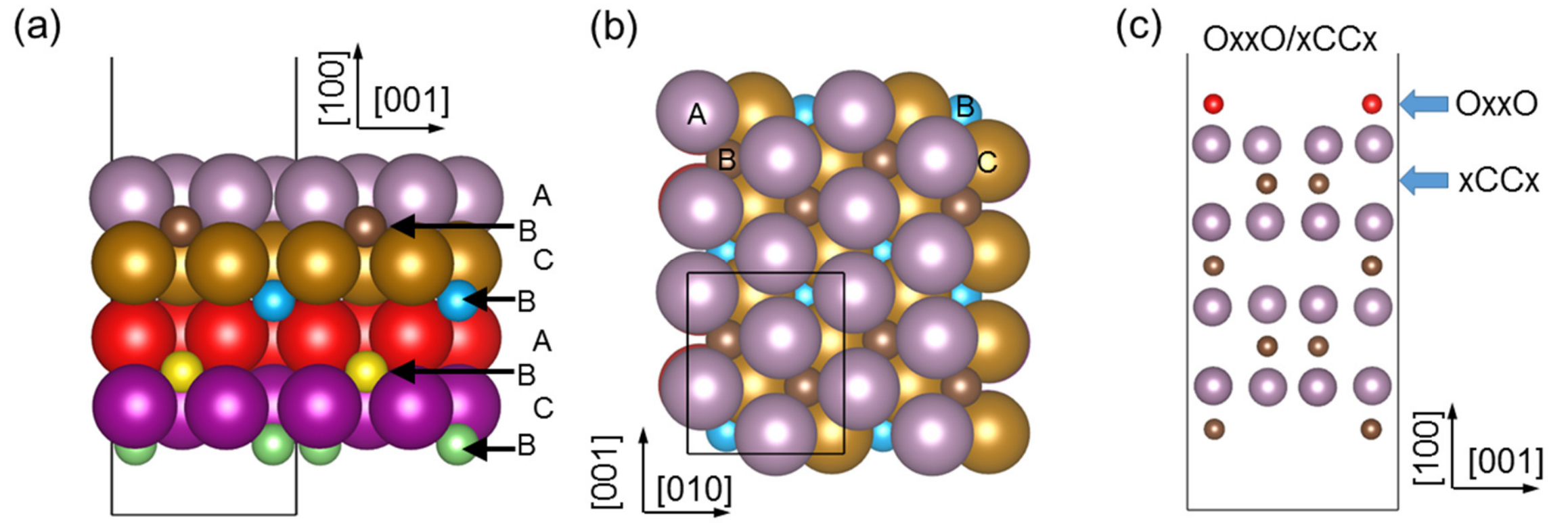

**Figure 1.** Side (a) and top (b) views of the Mo-terminated $Mo_2C(100)$ surface, with the layer stacking indicated. Larger atoms are Mo, smaller atoms are C, and each layer is assigned a different color. The convention used for the lattice constants and Miller indices of orthorhombic $Mo_2C$ is $a < b < c$. (c) Example of nomenclature of configurations of O and C atoms above and below the top Mo layer; the layer above the surface is of type *C*, while the layer below is *B* type.

## 2.b. Surface and subsurface configurations with deposited C and O atoms

In order to classify the configurations with O and C atoms above and below the surface, we start by describing the unit cell shown in Fig. 1a as a stacking sequence. Borrowing the well-known *ABC* nomenclature from face-centered cubic (fcc) structures, the stacking in orthorhombic $Mo_2C$ ($a < b < c$) is $ABCB\ ABCB\ \ldots$, where $A$ and $C$ are complete Mo layers (Fig. 1a, b). Each $B$ layer is only half-filled with carbon atoms, and there are two $B$ layers per bulk unit cell. To keep the calculations tractable, we use only $C$ sites above $A$ (top Mo layer) and only $B$ sites below the $A$ layer when adding O and C adsorbates. This choice is determined by the need to reproduce the ground state when only two carbons are in the second layer, and by the convenience of automatically avoiding $B$ site atoms in registry with each other, above and below the Mo $A$ layer. While such choice does not guarantee the global minimum energy minima, it does offer a systematic and representative set of surface and subsurface configurations characterized by specific O and C coverages. There are at most four $C$ sites per unit cell above and four $B$ sites below the surface that can be occupied by O and/or C atoms. To systematically enumerate the possibilities for occupying these $C$ and $B$ sites, we devised a notation (Fig. 1c) that lists the types of atoms (O, C, or x, where x indicates a vacant hollow site) above and below the surface and their position in a projection view along the [010] direction. For example, a structure denoted as OxxO/xCCx, indicates that above the Mo layer there are two O atoms in specific $C$ layer positions (first and fourth), and below the Mo layer there are two carbon atoms in the second and third positions (Fig. 1c), resulting in an oxygen surface coverage of 0.5 ML. We considered all the ($n_C$, $n_O$) ordered pairs with maximum four O and four C atoms ($0 \leq n_C \leq 4,\ 0 \leq n_O \leq 4$), which resulted in a total of 5407 structures. In addition, we considered several surface configurations with high oxygen content: ($n_C$ = 2, $n_O$ = 5), ($n_C$ = 2, $n_O$ = 6), ($n_C$ = 1, $n_O$ = 6), ($n_C$ = 1, $n_O$ = 7), and ($n_C$ = 0, $n_O$ = 8).

**2.c. Polymorph sampling for bulk oxycarbides**

The structure determination for bulk oxycarbides was carried out using a polymorph sampling method based on random superlattices (RSL),[66] in which initial random structures are generated so as to favor cation-anion coordinations from the outset. For each stoichiometry of the Mo-C-O system, we generated 1000 RSL structures. The minimum initial distance between two atoms was chosen to be 2.01 Å, which is the average cation-anion distance between the ground state structures of $MoO_2$ and $Mo_2C$. The RSL structures generated at each stoichiometry were relaxed (cell shape, volume, and atomic positions) using DFT.[62-64] In these DFT calculations, plane-wave energy cutoffs of 540 eV, and energy tolerances of $10^{-6}$ and $10^{-5}$ for electronic and ionic convergence were applied, respectively. The Brillouin zone was sampled by an automatically generated K point mesh ("Auto" function). Setting up workflow, performing DFT relaxations, and analysis of the results were handled through the pylada software.[67] While future studies may address refinements of the present low energy polymorphs via global optimization[68-71] or heuristic enumeration[72] methods, for our current purposes it is sufficient to have representative configurations at each composition investigated, i.e. structures characterized by low energy and highest frequency of occurrence amongst the relaxed RSL's.

## 3. Results

### 3.1 Stability of surface and subsurface oxygen and carbon

To determine the stability of surface and subsurface oxygen and carbon on $Mo_2C(100)$, we expanded on the work of Medford et al.,[41] in which surface oxidation was modeled by incorporation of O or C atoms in a few selected configurations either above or below the top Mo layer. We surmise that a significantly larger number of surface and subsurface configurations of O and C atoms compete in terms of thermodynamic stability. Therefore, we explored the relative stability of numerous possible configurations (described in detail in Sec. 2.b), and then identified those which can occur in a wide range of thermodynamic conditions.

The relative thermodynamic stability of these structures is dictated by their surface energy. For a slab with $n_{\mathrm{Mo_2C}}$ formula units in the supercell, the surface energy $\gamma$ is defined as the excess energy (per area) with respect to the bulk structure that is due to the presence of the surface and adsorbates,[41]

$$\gamma = \frac{1}{A}\left(E - n_{\mathrm{Mo_2C}}\mu_{\mathrm{Mo_2C}} - n_{\mathrm{O}}\mu_{\mathrm{O}} - \Delta n_{\mathrm{C}}\mu_{\mathrm{C}}\right) - \gamma_{\mathrm{b}}, \quad (1)$$

where $E$ is the total energy of the slab of surface area $A$, $\mu_{\mathrm{Mo_2C}}$ is the bulk structure chemical potential of the formula unit, $\mu_{\mathrm{O}}$ is the chemical potential of oxygen environment providing O adsorbates, and $n_{\mathrm{O}}$ is the number of oxygen atoms in the structure. Similarly, $\mu_{\mathrm{C}}$ is the chemical potential of carbon and $\Delta n_{\mathrm{C}}$ is the variation of the number of C atoms with respect to the original structure. The undetermined quantity $\gamma_{\mathrm{b}}$ in Eqn. (1) is the surface energy of the bottom face of the slab, a constant that does not influence the comparison between different surface energies. The term $n_{\mathrm{Mo_2C}}\mu_{\mathrm{Mo_2C}}/A$ is also constant across all surface structures. Strictly speaking, Eqn. (1) is an approximation since it does not include vibrational contributions; similar approximations are routinely used for reasonable assessments of surface stability,[54,73-74,] with the advantage of avoiding vibrational frequency calculations when dealing with a very large numbers of structures. While a deeper study of vibrational contributions would be warranted in the future, we note for now that such contributions to the surface energy were calculated by Wang et al.[74] to be 0.1-0.25 J/m$^2$ when the surface energies of various facets of orthorhombic $Mo_2C$ ranged from 1.2 to 5.4 J/m$^2$.

In order to assess the stability of surface and subsurface configurations, we compared the surface energies of the structures described above, including those with high oxygen content. The result of surface energy comparisons is the phase diagram in Fig. 2, which identifies the lowest surface energy structures (top views in the Supporting Information, SI, Figs. SI.1-SI.4) for selected intervals of $\mu_C$ and $\mu_O$. The phase diagram in Fig. 2 is in general agreement with previous results,[41] despite some quantitative differences due to variations of the PBE version and computational approximations. Our sequence of phases at the lower ends of the $\mu_O$ and $\mu_C$ ranges is the same as in Ref. 41, i.e. Mo, $Mo_4C$, $Mo_2C$, MoC from left to right on the bottom of Fig 2, and 0.25ML ($Mo_4O$), 0.5ML ($Mo_2O$), 0.75ML ($Mo_4O_3$) and 1ML (MoO) from bottom to top on the left of Fig. 2. Our phase diagram differs from previous reports in two key aspects: (i) surface and subsurface configurations were considered by Medford et al.[41] separately, and (ii) the large number of competing configurations in the present work results in a more complex phase diagram that includes phases which have not previously been reported.

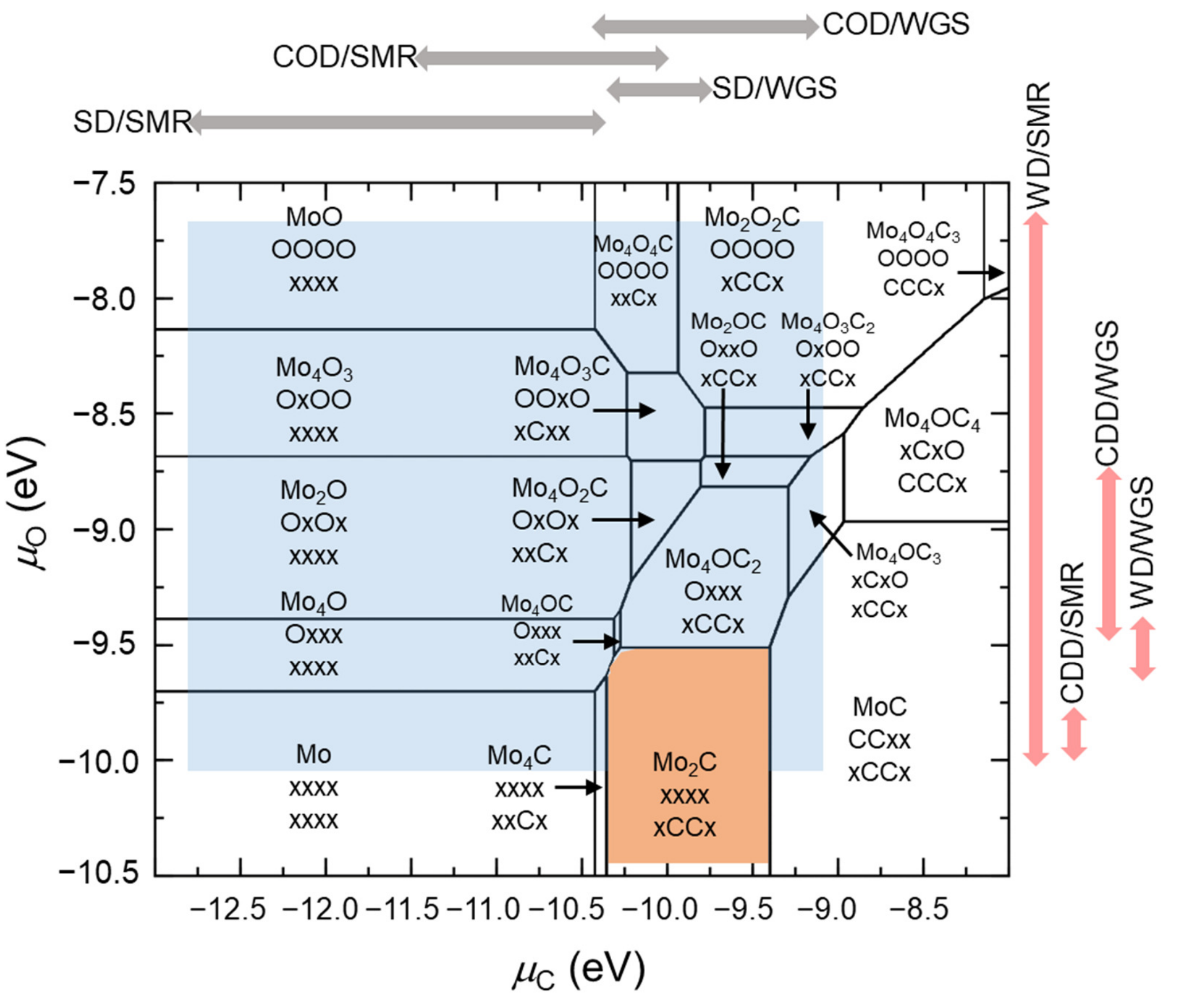

**Figure 2.** Phase diagram showing the surface configurations for $0 \leq n_C \leq 4,\ 0 \leq n_O \leq 4$. The stoichiometric notation of the top three atomic layers and short-hand notation of the surface and subsurface locations of O and C atoms. The block arrows on the top and right axes indicate the chemical potential ranges of C and O, respectively, under typical steam methane reforming (SMR) and water gas shift (WGS) reaction conditions (Tables SI.2, SI.3). The domain spanned by the chemical potentials is shown by the light blue rectangle, while the parent phase $Mo_2C$ is indicated in orange.

With such a diagram in place, it is useful to determine the ranges of chemical potentials of C and O that are relevant for ex situ CFP. The ranges of $\mu_C$ and $\mu_O$ depend on the reaction conditions (temperature, pressure, gas composition) for each of the possible deposition reactions. Determining the relevant values under ex situ CFP conditions is complicated by the breadth of reactants that supply C or O. To illustrate typical variations of the O and C chemical potentials, we calculated their ranges for realistic deposition reactions[75] under water-gas shift (WGS, $CO + H_2O \leftrightarrow H_2 + CO_2$) and steam methane reforming (SMR, $CH_4 + H_2O \leftrightarrow CO + 3H_2$) conditions. The deposition reactions[75] considered are carbon deposition via syngas deposition (SD, $CO + H_2$

↔ $H_2O$ + C*) or CO disproportionation (COD, 2CO ↔ $CO_2$ + C*), and oxygen deposition via water decomposition (WD, $H_2O$ ↔ $H_2$ + O*) or $CO_2$ decomposition (CDD, $CO_2$ ↔ CO + O*). A detailed description of the calculations and reaction conditions is included in SI.

The predicted oxygen surface coverage varies between 0 and 1 ML under SMR conditions when considering O deposition via the WD reaction. CDD is, however, insufficient to lead to oxygen deposition under SMR conditions, whereas both CDD and WD can lead to at least 25% oxygen coverage under WGS conditions. Interestingly, carbon deposition via either SD or COD under SMR reaction conditions can lead to the *removal* of carbon relative to the parent phase, as the chemical potential of C extends to the left of the parent phase region (Fig. 2). Carbon deposition via CO disproportionation (COD) under WGS conditions generates a carbon surface coverage of up to 50%, whereas carbon deposition via syngas deposition under WGS conditions results in a carbon content close to that of the parent phase. These results are consistent with those in prior reports,[6,41] and support experimental and theoretical evidence of the presence of a significant oxygen content in $Mo_2C$ under a wide range of reaction conditions. An important conclusion emerging from the phase diagram (Fig. 2) is that configurations of $Mo_2C$(100) with subsurface O are thermodynamically unstable at all reaction conditions, as no subsurface configurations make it in the diagram.

### 3.2 Surfaces with high oxygen content

However, there are indications that oxygen could populate sites below the top Mo layer.[76] To rationalize the presence of oxygen below the surface, we note that by virtue of their convex geometry, $Mo_2C$ catalyst nanoparticles cannot be exclusively bound by close-packed Mo layers, and thus it is conceivable that O atoms intercalate between the Mo planes and diffuse laterally between the layers. Results from the high oxygen content calculations ($n_O \geq 5$) indicate that the subsurface oxygen atoms (i.e., beneath the Mo $A$ layer) push the Mo layer upward, increasing the Mo-Mo interplanar distance. At $n_O = 8$ (the equivalent of 2 ML oxygen), the Mo-Mo interplanar distance increases from 2.35 to 2.94 Å. Concurrently, the triangular networks for the $C$ (O atoms), $A$ (Mo atoms), and $B$ (O atoms) layers distort significantly during the DFT relaxations (Fig. 3). The fixed surface unit vectors contribute to these distortions, suggesting scrutiny of the conclusions based on surface calculations. Indeed, even though they provide insight into the O and C content for various thermodynamic conditions, there are general caveats of the surface calculations. For example, the coverage values studied are limited by the surface unit cell used. In our case and others,[6,41] one unit cell (Fig. 1) allows only increments of 0.25 ML, resulting in missed configurations with intermediate values for the O or C surface content. Therefore, high oxygen content is better addressed in the context of bulk (as opposed to surface) structural changes, in which we allow for changes both in lattice vectors and in atomic positions. Bulk structural changes brought about by oxygen content are described below.

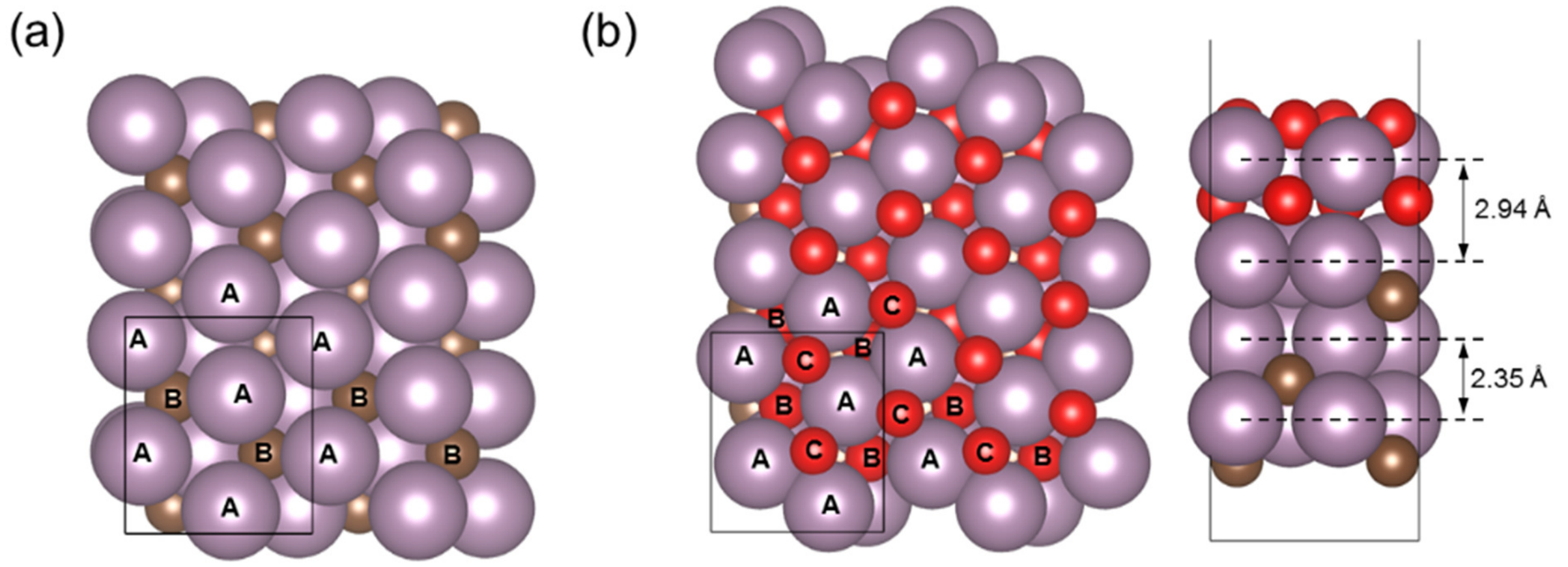

**Figure 3.** (a) Top view of the $Mo_2C(100)$ surface and (b) top and side views of a configuration with high oxygen content: 1 ML above the surface (*C* layer) and 1 ML below the surface (*B* layer). Mo, C, and O atoms are represented as large gray, small black, and small red spheres, respectively. The distortion of the triangular network for the OOOO/OOOO configuration is apparent in all neighboring layers, as well as the increase in the Mo-Mo interplanar spacing.

### 3.3. Predicted bulk $(Mo_2C)_xO_y$ oxycarbides

To assess bulk phase stability, we first determine the phases to which the orthorhombic phase could transform. Typical reactions (refer to Sec. 3.1 and SI) involve changing the oxygen content, the carbon content, or both. For clarity, we will focus here on changing only the oxygen content and consider stoichiometries with the Mo:C ratio of 2:1. To date, there are no known bulk oxycarbide $(Mo_2C)_xO_y$ structures, as gleaned, for example, via direct inspection of the Materials Project database.[77] We performed structural sampling for low-energy polymorphs of $(Mo_2C)_xO_y$ at twelve O concentrations, ranging from $y/x = 1/3$ to 4, with denser steps below $y/x = 2$. Ideally, a more comprehensive investigation into oxycarbide structures would include variations of the C content as well; however, for computational convenience, we restrict ourselves to scanning only the oxygen content. In terms of applicability of our results, this restriction means that the formation of oxycarbides should occur by varying *only* the oxygen chemical potential. This is strictly achieved when the only reaction present is the dissociative adsorption of molecular oxygen, which involves no changes in the C and Mo chemical potentials or in the Mo:C ratio. This is the case of the direct oxygen exposure studies.[46-47] Some of the structures presented here could occur under different reactions as well, provided that the Mo:C ratio does not change drastically.

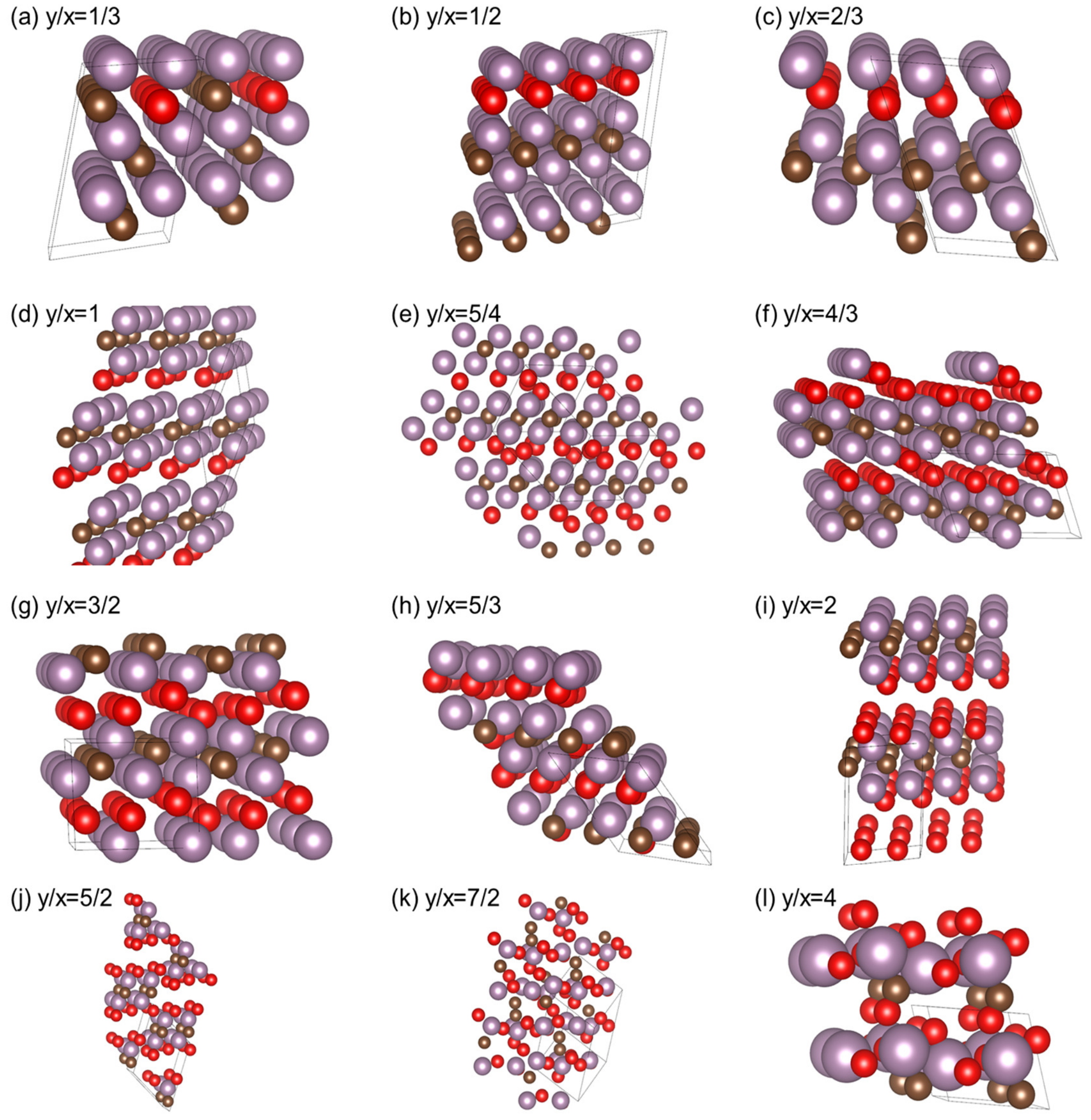

**Figure 4.** (a-l) Representative low energy $(Mo_2C)_xO_y$ oxycarbide structures determined from the polymorph sampling. Mo, C, and O atoms are represented as large gray, small black, and small red spheres, respectively. The gray lines delineate the bulk unit cells. Each figure is oriented so as to show the Mo planes (a-i). The Mo planes are rumpled and have interspersed O atoms for structures (j), (k), and (l).

The representative low energy structures obtained for each oxygen content are shown in Fig. 4, and the corresponding space group, lattice parameters, and lattice angles are listed in Table 1. The orientation and unit cell repetition for each individual panel in Fig. 4 have been chosen such that the Mo planes, when present, are obvious to the viewer. The Mo atomic planes are intact for $y/x \leq 2$ (Figs. 4a-i), whereas at higher concentrations the planes are either incomplete (i.e., uneven platelets, Figs. 4j, k) or atomically corrugated (Fig. 4l). We note that the oxygen and carbon atoms also form planes at concentrations $y/x \leq 1$ (Figs. 4a-d).

The Mo interplanar distances for each of the minimum energy structures in Fig. 4 are listed in Table 2, along with the average coordination numbers for Mo atoms determined using cutoff distances of 3.2 Å for Mo neighbors, and 2.4 Å for C and O neighbors. As the oxygen coordination ($N_{O}$) increases, the Mo coordination ($N_{Mo}$) decreases, until the Mo atoms no longer form atomic planes ($y/x > 2$).

**Table 1.** Space groups, lattice parameters, and lattice angles for orthorhombic $Mo_2C$ and the low energy $(Mo_2C)_xO_y$ oxycarbide structures determined via polymorph sampling. The oxycarbide structures are shown in Fig. 4.

| $y/x$ | Space group | $a$ (Å) | $b$ (Å) | $c$ (Å) | $\alpha$ (°) | $\beta$ (°) | $\gamma$ (°) |
|---|---|---|---|---|---|---|---|
| 0 | *Pbcn* | 4.789 | 5.251 | 6.073 | 90.00 | 90.00 | 90.00 |
| 1/3 | *Pm* | 2.958 | 7.819 | 6.006 | 79.23 | 60.49 | 90.00 |
| 1/2 | *R*3*m* | 3.003 | 11.207 | 3.003 | 105.57 | 120.00 | 90.00 |
| 2/3 | *Pm* | 3.063 | 5.346 | 8.699 | 66.54 | 90.00 | 90.00 |
| 1 | *R*3*m* | 7.646 | 6.459 | 2.892 | 89.96 | 124.58 | 139.49 |
| 5/4 | *P*1 | 6.289 | 6.375 | 7.769 | 107.87 | 128.47 | 103.33 |
| 4/3 | *Cm* | 8.589 | 6.295 | 2.990 | 103.75 | 89.98 | 67.85 |
| 3/2 | *Pm* | 6.302 | 3.028 | 5.187 | 90.00 | 92.55 | 90.01 |
| 5/3 | *P*1 | 5.995 | 6.788 | 6.081 | 98.08 | 60.46 | 61.69 |
| 2 | *P*1 | 5.875 | 2.938 | 8.135 | 93.33 | 93.36 | 59.31 |
| 5/2 | *C*2/*m* | 8.828 | 6.472 | 3.068 | 103.67 | 79.96 | 61.41 |
| 7/2 | *C*2 | 8.438 | 6.221 | 6.105 | 86.41 | 132.63 | 65.10 |
| 4.0 | *P*1 | 3.737 | 5.225 | 5.589 | 73.24 | 109.62 | 88.79 |

**Table 2.** Interplanar distances (*d*) for Mo planes, and average number of Mo ($N_{Mo}$), C ($N_C$), and O ($N_O$) neighbors for Mo atoms in $Mo_2C$ (first row) and in the $(Mo_2C)_xO_y$ oxycarbide phases reported in Fig. 4.

| $y/x$ | Figure | $d$ (Å) | $N_{Mo}$ | $N_C$ | $N_O$ |
|---|---|---|---|---|---|
| 0 | | 2.4 | 12 | 3 | 0 |
| 1/3 | 4a | 2.3, 3.0 | 10 | 3 | 0.833 |
| ½ | 4b | 2.3, 2.4, 3.4 | 10.5 | 3 | 1 |
| 2/3 | 4c | 2.4, 3.4 | 9.33 | 3 | 1.33 |
| 1 | 4d | 2.8, 3.5 | 7 | 3 | 2 |
| 5/4 | 4e | 2.1, 3.4 | 6 | 3 | 2.25 |
| 4/3 | 4f | 2.1, 3.6 | 6.67 | 3 | 2.5 |
| 3/2 | 4g | 2.1, 3.1 | 7 | 3 | 2.75 |
| 5/3 | 4h | 2.0, 3.2 | 5 | 2.33 | 3.33 |
| 2 | 4i | 2.8, 5.3 | 6.5 | 3 | 3 |
| 5/2 | 4j | – | 6 | 2.5 | 3.5 |
| 7/2 | 4k | – | 1 | 1 | 4 |
| 4 | 4l | – | 1 | 1 | 4.5 |

For any $y > 0$, we note that the distribution of Mo-plane spacings becomes approximately bimodal, with two nominal values (e.g., 2.8 and 3.6 Å for $y/x = 1$) instead of one (2.4 Å in the orthorhombic carbide phase). This is strictly due to O incorporation between atomic planes of Mo, which leads to wider Mo interlayer separations than the nominal $d$ in the parent phase. Furthermore, as depicted in Fig. 4, the structures of $(Mo_2C)_xO_y$ oxycarbides are not realized simply by inserting O atoms between the Mo planes; there are also changes in the stacking of the planes such that the $ABCB$ sequence (Fig. 1a) is not preserved and the crystal structure is no longer orthorhombic. In general, with increasing O content in the unit cell, the structures lose symmetry (Table 1). Based on the loss of atomic planarity for Mo, we propose that the structures with $y/x > 2$ indicate a tendency of amorphization in oxycarbides with high O content. This finding is consistent with a report wherein amorphous oxycarbides form mixtures with the parent $Mo_2C$ phase.[47]

The presence of Mo atomic planes for $y/x \leq 2$ and associated changes in the Mo-Mo interplanar distance with respect to the orthorhombic phase lead to direct signatures in XRD patterns, and hence the structures predicted here could be experimentally verified. To facilitate future comparisons with experiments, we simulated XRD patterns for the predicted structures using the Rietan software.[78] The simulated patterns (Fig. 5), which correspond to the Cu Kα wavelength, show marked differences between the oxycarbide phases (blue peaks) and the parent phase (red peaks). There is significant peak splitting in the oxycarbides for angles $2\theta$ larger than 30°, consistent with the variations in interplanar distances for the Mo planes due to oxygen incorporation (Table 2). This result was verified by additional XRD simulations of the bulk unit cells containing only the Mo atoms.

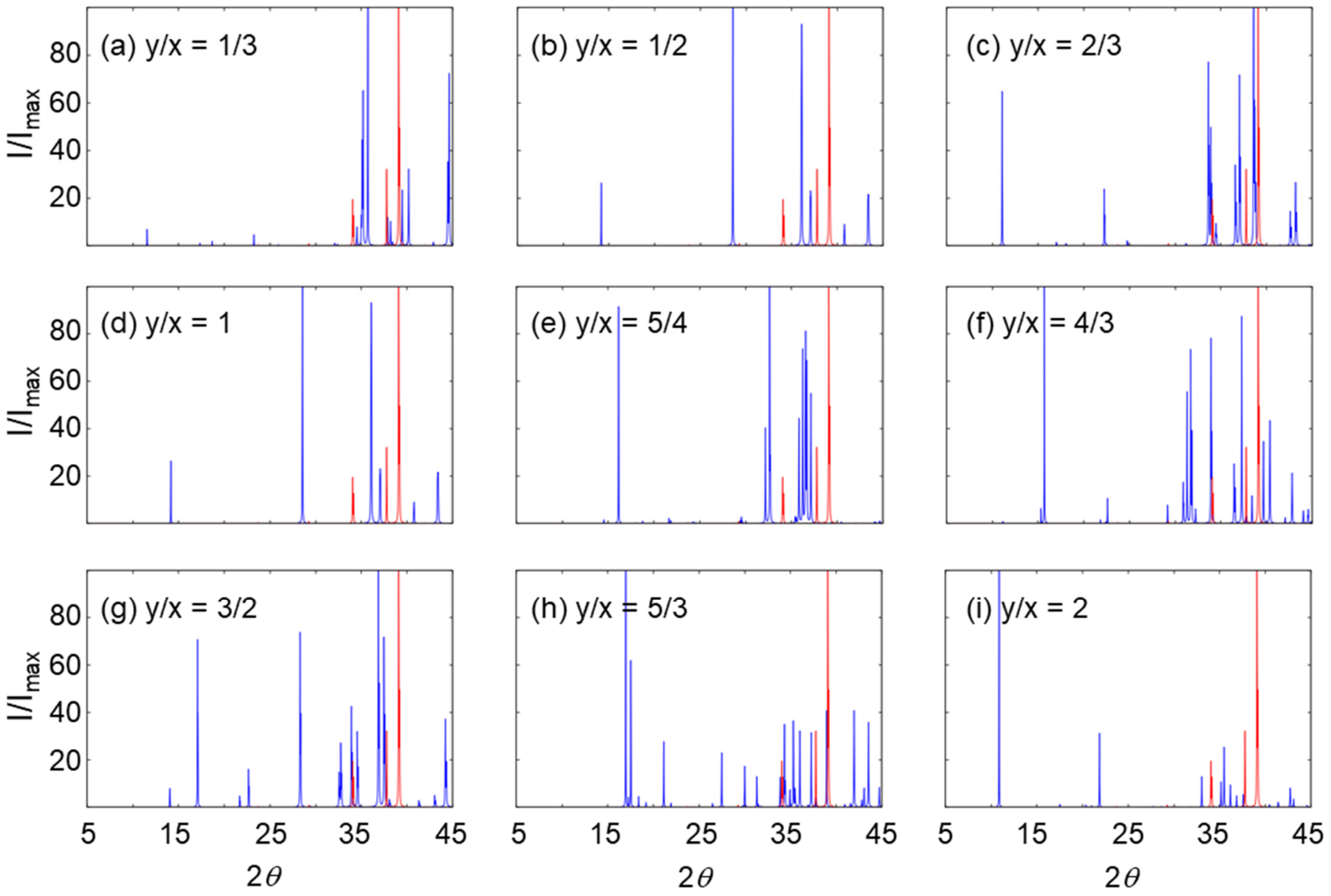

**Figure 5.** (a-l) Simulated XRD patterns (Cu Kα wavelength) for the $(Mo_2C)_xO_y$ oxycarbides predicted for $y/x \leq 2$. The XRD pattern of the parent $Mo_2C$ phase is shown in red in each frame.

The increase in the average coordination of Mo atoms with oxygen (Table 2) should also result in changes in the oxidation state of the Mo. To verify this, we calculated the net charge on Mo ions using Bader charge analysis.[79-80] Figure 6 shows the computed net charge for all Mo ions of the oxycarbides depicted in Fig. 4. As expected, the net charge (and hence the oxidation state) of Mo increases with increasing O content (Figure 6). We also note that the oxidation state differs across the Mo atoms within any oxycarbide unit cell, which is due to their different coordinations with Mo, C, and O. Somewhat qualitatively, the net charge on each Mo can be converted to an approximate oxidation state or valence by scaling the charge on all atoms so that the O ions have a $-2$ charge, on average (Fig. SI. 5). The valence of Mo in oxycarbides disperses over an interval, rather than assuming fixed values. For compositions with $y/x < 1$, the Mo oxidation states are between 1 and 2.7 (Fig. SI. 5), whereas for compositions with $y/x > 1$ the Mo oxidation states are between +2 and +4. Most of these states are distinct from the oxidation state of Mo in the parent $Mo_2C$ phase (+2) and should appear in experimental XPS spectra. Mo (3d) peaks in XPS studies are often assigned to oxidation states between +3 and +6, although it is difficult to determine accurately whether these species are due to an oxycarbide phase or to coexisting Mo oxide phases.[81-83]

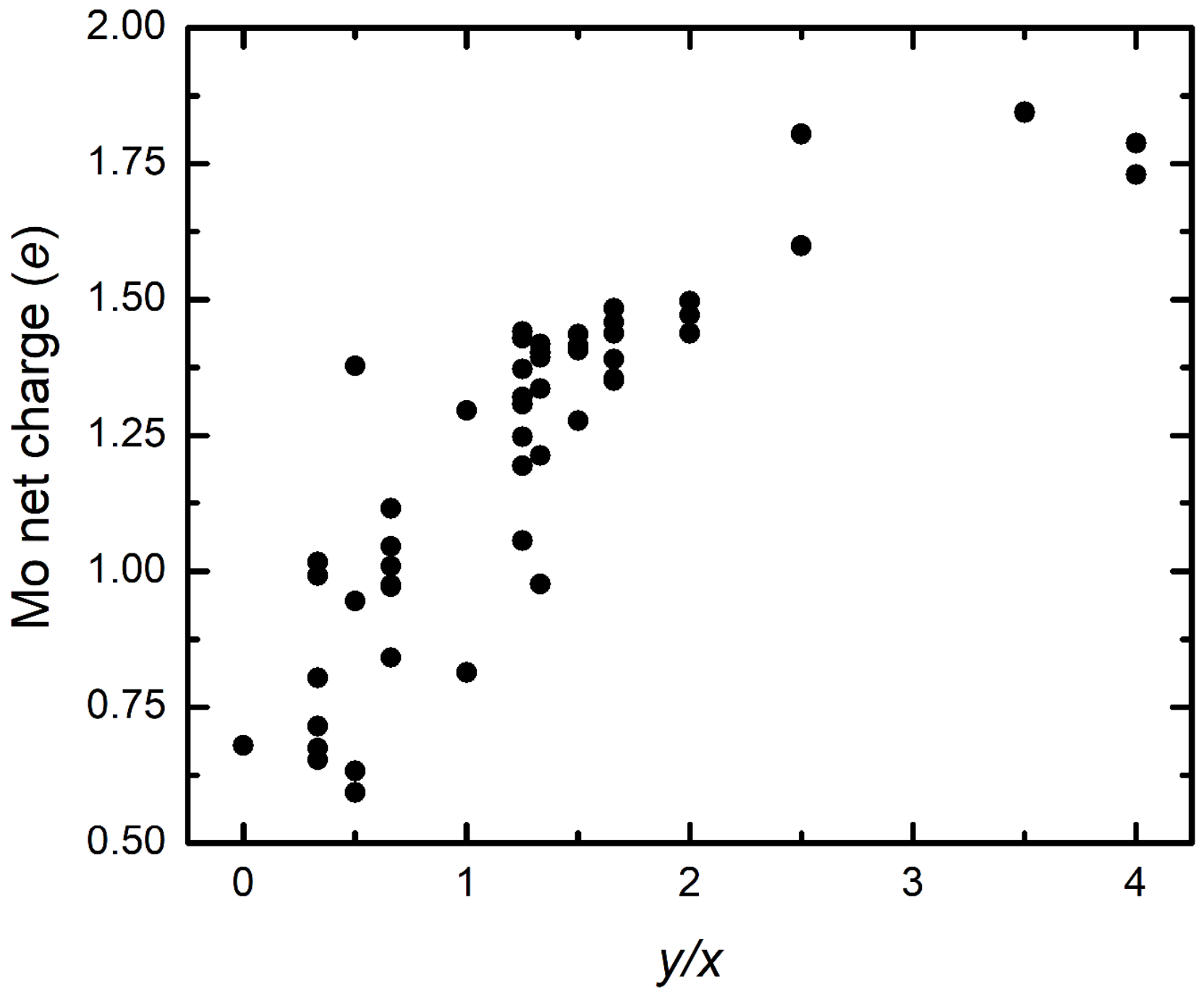

**Figure 6.** Net charge on Mo ions from Bader charge analysis, for each of the low-energy oxycarbide phases obtained from polymorph sampling.

### 3.4. Stability of $(Mo_2C)_xO_y$ oxycarbides

The structures obtained from the bulk $(Mo_2C)_xO_y$ oxycarbide polymorph sampling reported here could be the subject of future experimental investigations, provided they are sufficiently stable. Similar to the case of the surface configurations, we can assess the relative thermodynamic stability of bulk $(Mo_2C)_xO_y$ oxycarbides by comparing their grand potential per volume ($g$) when the number of Mo, C, and O vary (independently, in general):

$$g = \frac{1}{V}\left(E - n_{\mathrm{Mo}}\,\mu_{\mathrm{Mo}} - n_{\mathrm{C}}\mu_{\mathrm{C}} - n_{\mathrm{O}}\mu_{\mathrm{O}}\right), \quad (2)$$

where $E$and $V$ are the total energy and volume of the bulk unit cell, respectively, and in which entropic contributions are neglected. Equation (2) is akin to Eqn. (1), while illustrating that the phase transformation(s) would occur per volume (rather than area) and that the number of Mo atoms may vary. For our case of fixed Mo:C ratio, the chemical potentials of Mo and C are related through $\mu_{\mathrm{Mo_2C}} = 2\mu_{\mathrm{Mo}} + \mu_{\mathrm{C}}$ , which changes Eqn. (2) to

$$g = \frac{1}{V}\left(E - n_{\mathrm{Mo_2C}}\,\mu_{\mathrm{Mo_2C}} - n_{\mathrm{O}}\mu_{\mathrm{O}}\right), \quad (3)$$

Equation (3) describes the formation energy per volume of oxycarbide, formed (effectively) from $Mo_2C$ units provided by the parent carbide film or nanoparticles acting as a reservoir with chemical potential $\mu_{\mathrm{Mo_2C}}$, and from oxygen provided via certain deposition reactions ($\mu_{\mathrm{O}}$). Positive $g$ values indicate that oxycarbide formation is endothermic, whereas negative values indicate that oxycarbide formation should be thermodynamically favorable (exothermic).

We have plotted Eqn. (3) as a function of the chemical potential $\mu_{\mathrm{O}}$ in Fig. 7. The $Mo_2C$ orthorhombic phase is more stable than any $(Mo_2C)_xO_y$ oxycarbide for all $\mu_{\mathrm{O}}$ values smaller than approximately $-7.25$ eV (Fig. 7). This value of $\mu_{\mathrm{O}}$ is outside the chemical potential range that can be reached under SMR or WGS reaction conditions ($-10.01 \leq \mu_{\mathrm{O}} \leq -7.67$ eV, Fig. 2 and Table SI.3). This indicates that bulk phase transitions to $(Mo_2C)_xO_y$ oxycarbides are not thermodynamically favorable under these reaction conditions: instead, SMR and WGS lead only to the deposition of oxygen on the surface (Fig. 2). However, direct dissociation of molecular oxygen ($O_2 \leftrightarrow 2O^*$) can result in a higher chemical potential ($\mu_{\mathrm{O}} \leq -6.36$ eV) under WGS conditions (473 K – 723K, assuming a 0.01−0.4 bar $O_2$ partial pressure range). Such increase in $\mu_{\mathrm{O}}$ would trigger the transformation to $(Mo_2C)_xO_y$ oxycarbides (refer to Fig. 7). In this case,

$(Mo_2C)_xO_y$ oxycarbides

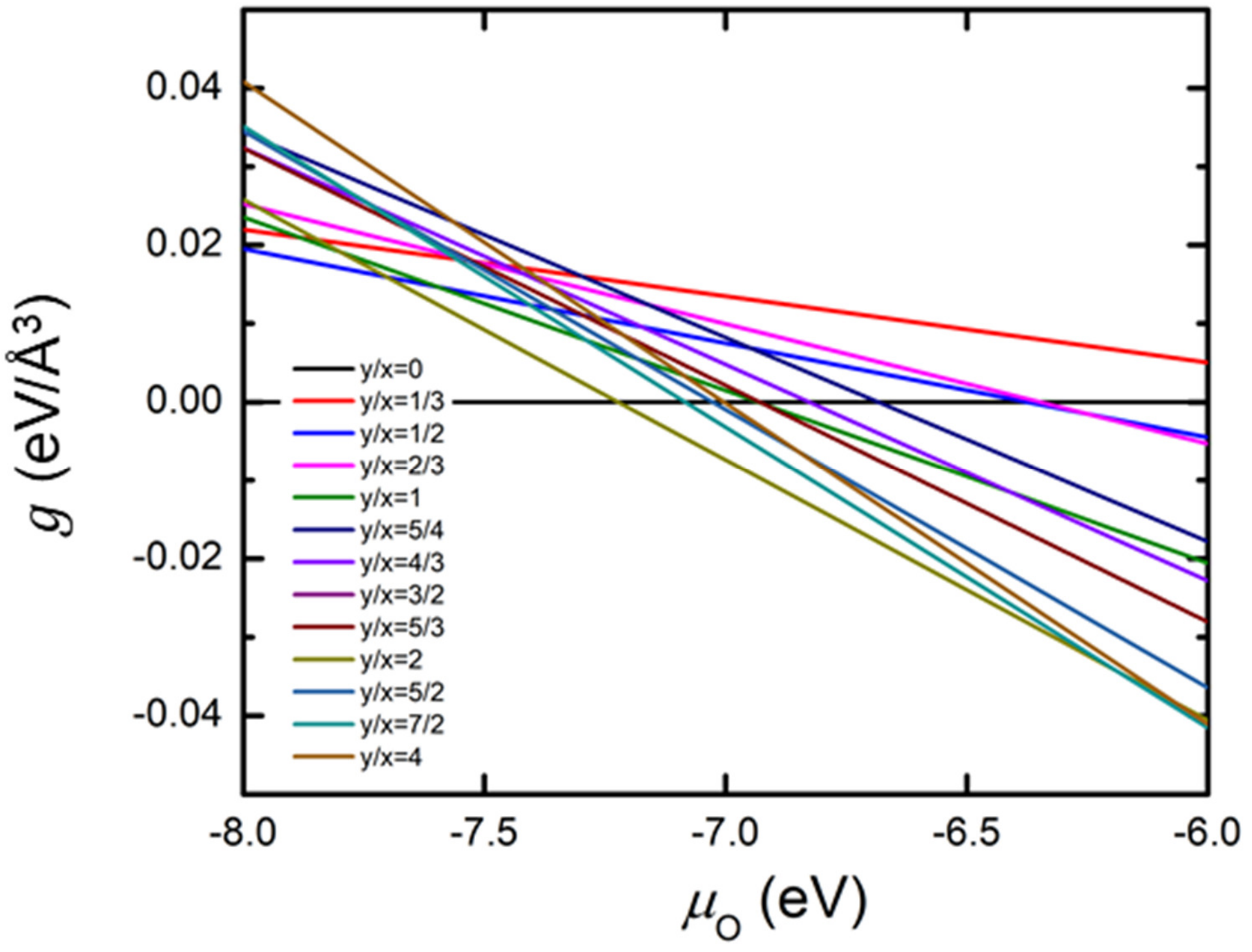

**Figure 7.** Stability of $(Mo_2C)_xO_y$ bulk phases assessed through the grand potential per volume.

become more stable than the parent carbide at a high oxygen content ($y/x = 2$, Fig. 7), rather than gradually. At $y/x > 2$ and sufficiently high $\mu_O$, the nanoparticles may even adopt amorphous structures. In general, we point out that there could also be avenues for bulk phase transformations which involve variation of the Mo:C ratio. This was pointed out in the literature,[46] and is particularly relevant for ex situ CFP where the effects of varying the chemical potentials of O and C are inter-related. For example, $\mu_C$ for syngas deposition under SMR conditions leads to the depletion of C atoms from the parent carbide (Fig. 2 and Table SI.3), which would free up sites between Mo planes thereby enhancing the propensity for incorporating oxygen atoms. While the kinetics of oxygen intercalation between Mo planes is beyond the scope of this study, future work on oxygen diffusion and, separately, on the stability of oxycarbides in which *both* C and O content are allowed to vary could offer more insight into the formation of oxycarbides and their role in the degradation of carbide catalyst.

## 4. Discussion

In practice, there are two main situations in which Mo oxycarbides may form: (i) oxidation of $Mo_2C$ catalysts by either molecular oxygen or environments rich in oxygen-containing compounds such as $H_2O$, CO, $CO_2$, and organic oxygenates,[42,46,84] and (ii) incomplete carburization of Mo oxides.[47,51] As discussed above, the oxidation of $Mo_2C$ occurs at the surface[6,42] and is not thermodynamically favorable to proceed through bulk under SMR and WGS conditions (Fig. 7 and $\mu_O$range in Table SI.3). The presence of molecular oxygen will, however, lead to the formation of bulk oxycarbides since the oxygen dissociation reaction can increase the chemical potential of deposited oxygen past the carbide-oxycarbide transition point ($-7.25$ eV, Fig. 7).

We noted (Sec. 3.3) that the assignment of the Mo (3d) peaks in XPS to specific oxidation states of Mo is difficult due to the possible coexistence of the oxycarbides with $MoO_2$, $MoO_3$ or $Mo_2C$ during carburization process. The process of oxidation of the carbide, however, may prove to be less ambiguous in terms of such assignments if oxidation is performed controllably starting from the pure carbide;[46,76] this is because upon gradually increasing exposure to molecular oxygen the crystalline oxycarbide phase(s) would form *prior to* the $MoO_2$, $MoO_3$ or amorphous phases which are expected to form only at sufficiently high oxygen exposure. As such, Edamoto at al.[76] attribute some of the XPS peaks appeared after the oxygen exposure to Mo atoms coordinated with 2 or 3 oxygens: such coordinations O ($2 \leq N_O \leq 3$) can occur for $y/x$ from 1 to 2 (Table 2).

Óvári et al. carried out an oxidation study over a wide range of oxygen exposures, at temperatures from 300 to 1000K. Oxygen adsorbs dissociatively on the surface at room temperature, but leads to bulk oxidation of the carbide at temperatures above 800K; in the latter case, the O:Mo ratio increases, while the C:Mo ratio decreases with oxygen exposure.[46] It is not clear if the oxycarbides formed by oxidation at temperatures above 800K are crystalline,[46] but XPS characterization revealed an increase in the density of +4 Mo states, consistent with the oxidations states that computed here for $y/x > 2$ (Fig. 6 and Fig. SI.5 in which the valence of Mo atoms is estimated from scaled Bader charges).

Delporte et al.[51] carried out systematic studies of carburization of Mo oxides, and were able to identify the presence of one crystalline oxycarbide phase. $MoO_3$ exposed to a mixture of flowing n-cyclohexane and hydrogen resulted in the formation of a new phase with XRD peaks at interplanar distances of 2.01, 4.1, and 6.2 Å.[51] High-resolution transmission electron microscopy shows that this crystalline phase coexists with an amorphous phase also formed during carburization, and also with crystalline $MoO_2$.[51] These authors carried out elemental analysis,[52] which they corrected for the presence of $MoO_2$ using their XRD data. As a result, they report a formula of $MoO_{2.9}C_{0.8}$ for their oxycarbide, although we suspect that this could likely be an average stoichiometry between the crystalline form and the amorphous oxycarbide matrix in which it is embedded. We surmise that the amorphous oxycarbide has a large $y/x$ ratio, and compare our XRD simulations for $y/x < 2$ with the measurements of Delporte et al.[51] Upon conversion of the simulated XRD patterns to the Co wavelength, the only structure in Figure 4 that has a pattern resembling that in XRD experiments is that obtained for $y/x = 1$. $Mo_2CO$ ($y/x = 1$, Figure 4) matches within ~1° each of the three experimental peak locations $2\theta$ associated by Delporte et al. with the crystalline oxycarbide (their Figure 1d).[51] Our other simulated peaks narrowly coincide with experimental $MoO_2$ peaks, which therefore could not have been unambiguously assigned to the oxycarbide based only on experimental data. We suggest that the structure in Fig. 4d is related to the crystalline oxycarbide in experiments.[51-52] However, future studies seem warranted in order to fully elucidate the fraction of crystalline (vs. amorphous) oxycarbide in experiments,[51-52] its stoichiometry, and atomic structure.

## 5. Conclusions

In conclusion, we have analyzed the structure of surface and bulk oxycarbides and discussed their thermodynamic stability for different chemical potentials for carbon and oxygen. The surface structures were enumerated systematically based on occupying hollow sites with O and C above and below the top Mo layer, while the bulk oxycarbide structures were determined via a polymorph sampling method coupled with DFT calculations. To understand the relevance of these structures for ex situ CFP, we determined their relative stabilities over a large range of oxygen and carbon chemical potentials, including those representative of realistic reaction conditions encountered in SMR and WGS. We found that the bulk transformation to oxycarbides is not thermodynamically favorable; however, the SMR and WGS conditions may lead to changes in the C and O content of the layers above and below Mo. If molecular oxygen is present, then the oxygen dissociation reaction can lead to the formation of bulk oxycarbides as this reaction increases the chemical potential of deposited oxygen past the highest value achievable with SMR and WGS conditions (Table SI.3), and past the point at which oxycarbides become stable (Fig. 7).

In order to compare our results with experiments, we determined the changes in oxidation states of Mo, and have simulated the XRD patterns of the bulk oxycarbides predicted in this study. Increasing concentration of oxygen leads to increases in the (average) valence of Mo, which is consistent with experiments.[46] One of the oxycarbide structures identified in this study was found to exhibit a simulated XRD pattern very similar to experiments[51-52] and we suggest that the crystalline oxycarbide reported in literature to date has the structure and composition similar to that shown in Fig. 4d. Our study could serve as starting point for determining (i) phase coexistence in metal carbide/oxycarbide systems, (ii) synthesis conditions for oxycarbide catalysts, and (iii) the catalytic properties and exposed surfaces of crystalline oxycarbides over a range of O and C contents. In the future, building a composition-temperature phase diagram for the Mo-O-C system would serve as a guideline for designing oxycarbide catalysts.

**Supporting Information.** Supporting information for publication includes (i) surface configurations, (ii) deposition reactions, chemical potential calculations, and reaction conditions, and (iii) Mo valence calculations from scaled Bader charges.

**Acknowledgments.** The research at National Renewable Energy Laboratory (NREL) was supported by the U.S. Department of Energy Bioenergy Technologies Office, Contract No. DE-AC36-08-GO28308, in collaboration with the Chemical Catalysis for Bioenergy Consortium (ChemCatBio), a member of the Energy Materials Network (EMN). The research at Colorado School of Mines was supported through the Alliance Partner University Program of NREL through Contract No. UGA-0-41025-116, and also in part by the National Science Foundation through Grant. No. DMREF-1534503. The DFT calculations were performed at the high-performance computing centers of NREL and Colorado School of Mines (Golden Energy Computing Organization).

## References

(1) Ruddy, D. A.; Schaidle, J. A.; Ferrell, J. R.; Wang, J.; Moens, L.; Hensley, J. E. Recent Advances in Heterogeneous Catalysts for Bio-Oil Upgrading Via "Ex Situ Catalytic Fast Pyrolysis": Catalyst Development through the Study of Model Compounds. *Green Chem.* **2014,** *16*, 454−490.

(2) Bridgwater, A. V. Review of Fast Pyrolysis of Biomass and Product Upgrading. *Biomass Bioenergy* **2012,** *38*, 68−94.

(3) Carlson, T. R.; Tompsett, G. A.; Conner, W. C.; Huber, G. W. Aromatic Production from Catalytic Fast Pyrolysis of Biomass-Derived Feedstocks. *Top. Catal.* **2009,** *52*, 241−252.

(4) Alonso, D. M.; Bond, J. Q.; Dumesic, J. A. Catalytic Conversion of Biomass to Biofuels. *Green Chem.* **2010,** *12*, 1493−1513.

(5) Garron, A.; Al Maksoud, W.; Larabi, C.; Arquilliere, P.; Szeto, K. C.; Walter, J. J.; Santini, C. C. Direct Thermo-Catalytic Transformation of Pine Wood into Low Oxygenated Fuel: Influence of the Support. *Catal. Today* **2015,** *255*, 75−79.

(6) Schaidle, J. A.; Blackburn, J.; Farberow, C. A.; Nash, C.; Steirer, K. X.; Clark, J.; Robichaud, D. J.; Ruddy, D. A. Experimental and Computational Investigation of Acetic Acid Deoxygenation over Oxophilic Molybdenum Carbide: Surface Chemistry and Active Site Identity. *ACS Catal.* **2016,** *6*, 1181−1197.

(7) Choi, J. S.; Schwartz, V.; Santillan-Jimenez, E.; Crocker, M.; Lewis, S. A.; Lance, M. J.; Meyer, H. M.; More, K. L. Structural Evolution of Molybdenum Carbides in Hot Aqueous Environments and Impact on Low-Temperature Hydroprocessing of Acetic Acid. *Catalysts* **2015,** *5*, 406−423.

(8) Abraham, D.; Nagy, B.; Dobos, G.; Madarasz, J.; Onyestyak, G.; Trenikhin, M. V.; Laszlo, K. Hydroconversion of Acetic Acid over Carbon Aerogel Supported Molybdenum Catalyst. *Microporous Mesoporous Mater.* **2014,** *190*, 46−53.

(9) Nash, C. P.; Farberow, C. A.; Hensley, J. E. Temperature-Programmed Deoxygenation of Acetic Acid on Molybdenum Carbide Catalysts. *J. Vis. Exp.* **2017**, e55314.

(10) Ren, H.; Yu, W. T.; Salciccioli, M.; Chen, Y.; Huang, Y. L.; Xiong, K.; Vlachos, D. G.; Chen, J. G. G. Selective Hydrodeoxygenation of Biomass-Derived Oxygenates to Unsaturated Hydrocarbons Using Molybdenum Carbide Catalysts. *ChemSusChem* **2013,** *6*, 798−801.

(11) Jongerius, A. L.; Gosselink, R. W.; Dijkstra, J.; Bitter, J. H.; Bruijnincx, P. C. A.; Weckhuysen, B. M. Carbon Nanofiber Supported Transition-Metal Carbide Catalysts for the Hydrodeoxygenation of Guaiacol. *Chemcatchem* **2013,** *5*, 2964−2972.

(12) Chang, J.; Danuthai, T.; Dewiyanti, S.; Wang, C.; Borgna, A. Hydrodeoxygenation of Guaiacol over Carbon-Supported Metal Catalysts. *Chemcatchem* **2013,** *5*, 3041−3049.

(13) Mortensen, P. M.; de Carvalho, H. W. P.; Grunwaldt, J. D.; Jensen, P. A.; Jensen, A. D. Activity and Stability of Mo2c/Zro2 as Catalyst for Hydrodeoxygenation of Mixtures of Phenol and 1-Octanol. *J. Catal.* **2015,** *328*, 208−215.

(14) Santillan-Jimenez, E.; Perdu, M.; Pace, R.; Morgan, T.; Crocker, M. Activated Carbon, Carbon Nanofiber and Carbon Nanotube Supported Molybdenum Carbide Catalysts for the Hydrodeoxygenation of Guaiacol. *Catalysts* **2015,** *5*, 424−441.

(15) Engelhardt, J.; Lyu, P. B.; Nachtigall, P.; Schuth, F.; Garcia, A. M. The Influence of Water on the Performance of Molybdenum Carbide Catalysts in Hydrodeoxygenation Reactions: A Combined Theoretical and Experimental Study. *Chemcatchem* **2017,** *9*, 1985−1991.

(16) Levy, R. B.; Boudart, M. Platinum-Like Behavior of Tungsten Carbide in Surface Catalysis. *Science* **1973,** *181*, 547−549.

(17) Sinfelt, J. H.; Yates, D. J. Effect of Carbiding on the Hydrogenolysis Activity of Molybdenum. *Nature* **1971,** *229*, 27−28.

(18) Hwu, H. H.; Chen, J. G. G. Surface Chemistry of Transition Metal Carbides. *Chem. Rev.* **2005,** *105*, 185−212.
(19) Vidick, B.; Lemaitre, J.; Leclercq, L. Control of the Catalytic Activity of Tungsten Carbides .3. Activity for Ethylene Hydrogenation and Cyclohexane Dehydrogenation. *J. Catal.* **1986,** *99*, 439−448.
(20) Lee, J. S.; Yeom, M. H.; Park, K. Y.; Nam, I. S.; Chung, J. S.; Kim, Y. G.; Moon, S. H. Preparation and Benzene Hydrogenation Activity of Supported Molybdenum Carbide Catalysts. *J. Catal.* **1991,** *128*, 126−136.
(21) Djegamariadassou, G.; Boudart, M.; Bugli, G.; Sayag, C. Modification of the Surface-Composition of Molybdenum Oxynitride During Hydrocarbon Catalysis. *Catal. Lett.* **1995,** *31*, 411−420.
(22) Ranhotra, G. S.; Bell, A. T.; Reimer, J. A. Catalysis over Molybdenum Carbides and Nitrides .2. Studies of Co Hydrogenation and C2h6 Hydrogenolysis. *J. Catal.* **1987,** *108*, 40−49.
(23) Liu, J.; Shen, J.; Gao, X.; Lin, L. Thermogravimetric Study of the Carburization and Coking of Unsupported and Carbon-Supported Fe, Mo and Fe-Mo Catalysts for Fischer-Tropsch Synthesis. *J. Therm. Anal. Calorim.* **1993,** *40*, 1239−1244.
(24) York, A. P. E.; Claridge, J. B.; Marquez-Alvarez, C.; Brungs, A. J.; Tsang, S. C.; Green, M. L. H. In *3rd World Congress on Oxidation Catalysis*; Grasselli, R. K., Oyama, S. T., Gaffney, A. M., Lyons, J. E., Eds., 1997; Vol. 110.
(25) Aegerter, P. A.; Quigley, W. W. C.; Simpson, G. J.; Ziegler, D. D.; Logan, J. W.; McCrea, K. R.; Glazier, S.; Bussell, M. E. Thiophene Hydrodesulfurization over Alumina-Supported Molybdenum Carbide and Nitride Catalysts: Adsorption Sites, Catalytic Activities, and Nature of the Active Surface. *J. Catal.* **1996,** *164*, 109−121.
(26) Dhandapani, B.; St Clair, T.; Oyama, S. T. Simultaneous Hydrodesulfurization, Hydrodeoxygenation, and Hydrogenation with Molybdenum Carbide. *Appl. Catal. A* **1998,** *168*, 219−228.
(27) Liu, P.; Rodriguez, J. A.; Asakura, T.; Gomes, J.; Nakamura, K. Desulfurization Reactions on Ni2p(001) and Alpha-Mo2c(001) Surfaces: Complex Role of P and C Sites. *J. Phys. Chem. B* **2005,** *109*, 4575−4583.
(28) Garcia, A. M.; He, J. J.; Lyu, P. B.; Nachtigall, P. Exploring the Stability and Reactivity of Ni2p and Mo2c Catalysts Using Ab Initio Atomistic Thermodynamics and Conceptual Dft Approaches. *Biomass Convers. Biorefin.* **2017,** *7*, 377−383.
(29) Ledoux, M. J.; Hantzer, S.; Huu, C. P.; Guille, J.; Desaneaux, M. P. New Synthesis and Uses of High-Specific-Surface Si C as a Catalytic Support That Is Chemically Inert and Has High Thermal-Resistance. *J. Catal.* **1988,** *114*, 176−185.
(30) Leclercq, L.; Provost, M.; Pastor, H.; Leclercq, G. Catalytic Properties of Transition-Metal Carbides .2. Activity of Bulk Mixed Carbides of Molybdenum and Tungsten in Hydrocarbon Conversion. *J. Catal.* **1989,** *117*, 384−395.
(31) Ledoux, M. J.; Phamhuu, C. High Specific Surface-Area Carbides of Silicon and Transition-Metals for Catalysis. *Catal. Today* **1992,** *15*, 263−284.
(32) Chorley, R. W.; Lednor, P. W. Synthetic Routes to High Surface-Area Nonoxide Materials. *Adv. Mater.* **1991,** *3*, 474−485.
(33) Preiss, H.; Meyer, B.; Olschewski, C. Preparation of Molybdenum and Tungsten Carbides from Solution Derived Precursors. *J. Mater. Sci.* **1998,** *33*, 713−722.
(34) Xiao, T. C.; York, A. P. E.; Williams, V. C.; Al-Megren, H.; Hanif, A.; Zhou, X. Y.; Green, M. L. H. Preparation of Molybdenum Carbides Using Butane and Their Catalytic Performance. *Chem. Mater.* **2000,** *12*, 3896−3905.
(35) Guzman, H. J.; Xu, W. Q.; Stacchiola, D.; Vitale, G.; Scott, C. E.; Rodriguez, J. A.; Pereira-Almao, P. In Situ Time-Resolved X-Ray Diffraction Study of the Synthesis of Mo2c with Different Carburization Agents. *Can. J. Chem.* **2013,** *91*, 573−582.

(36) Xu, W. Q.; Ramirez, P. J.; Stacchiola, D.; Brito, J. L.; Rodriguez, J. A. The Carburization of Transition Metal Molybdates (Mxmoo4, M = Cu, Ni or Co) and the Generation of Highly Active Metal/Carbide Catalysts for Co2 Hydrogenation. *Catal. Lett.* **2015,** *145*, 1365−1373.

(37) Volpe, L.; Boudart, M. Compounds of Molybdenum and Tungsten with High Specific Surface-Area .2. Carbides. *J. Solid State Chem.* **1985,** *59*, 348−356.

(38) Abe, H.; Bell, A. T. Catalytic Hydrotreating of Indole, Benzothiophene, and Benzofuran over Mo2n. *Catal. Lett.* **1993,** *18*, 1−8.

(39) Claridge, J. B.; York, A. P. E.; Brungs, A. J.; Marquez-Alvarez, C.; Sloan, J.; Tsang, S. C.; Green, M. L. H. New Catalysts for the Conversion of Methane to Synthesis Gas: Molybdenum and Tungsten Carbide. *J. Catal.* **1998,** *180*, 85−100.

(40) Decker, S.; Lofberg, A.; Bastin, J. M.; Frennet, A. Study of the Preparation of Bulk Tungsten Carbide Catalysts with C2h6/H-2 and C2h4/H-2 Carburizing Mixtures. *Catal. Lett.* **1997,** *44*, 229−239.

(41) Medford, A. J.; Vojvodic, A.; Studt, F.; Abild-Pedersen, F.; Norskov, J. K. Elementary Steps of Syngas Reactions on Mo2c(001): Adsorption Thermochemistry and Bond Dissociation. *J. Catal.* **2012,** *290*, 108−117.

(42) Liu, P.; Rodriguez, J. A. Water-Gas-Shift Reaction on Molybdenum Carbide Surfaces: Essential Role of the Oxycarbide. *J. Phys. Chem. B* **2006,** *110*, 19418−19425.

(43) Shi, X. R.; Wang, S. G.; Wang, J. G. Chemisorption of Oxygen and Subsequent Reactions on Low Index Surfaces of Beta-Mo2c: Insights from First-Principles Thermodynamics and Kinetics. *J. Mol. Catal. A Chem.* **2016,** *417*, 53−63.

(44) Nagai, M.; Kurakami, T.; Omi, S. Activity of Carbided Molybdena-Alumina for Co2 Hydrogenation. *Catal. Today* **1998,** *45*, 235−239.

(45) Nagai, M.; Oshikawa, K.; Kurakami, T.; Miyao, T.; Omi, S. Surface Properties of Carbided Molybdena-Alumina and Its Activity for Co2 Hydrogenation. *J. Catal.* **1998,** *180*, 14−23.

(46) Ovari, L.; Kiss, J.; Farkas, A. P.; Solymosi, F. Surface and Subsurface Oxidation of Mo2c/Mo(100): Low-Energy Ion-Scattering, Auger Electron, Angle-Resolved X-Ray Photoelectron, and Mass Spectroscopy Studies. *J. Phys. Chem. B* **2005,** *109*, 4638−4645.

(47) Wang, H.; Liu, S.; Smith, K. J. Synthesis and Hydrodeoxygenation Activity of Carbon Supported Molybdenum Carbide and Oxycarbide Catalysts. *Energy Fuels* **2016,** *30*, 6039−6049.

(48) Hwu, H. H.; Zellner, M. B.; Chen, J. G. G. The Chemical and Electronic Properties of Oxygen-Modified C/Mo(110): A Model System for Molybdenum Oxycarbides. *J. Catal.* **2005,** *229*, 30−44.

(49) Phamhuu, C.; Ledoux, M. J.; Guille, J. Reactions of 2-Methylpentane and 3-Methylpentane, Methylcyclopentane, Cyclopentane, and Cyclohexane on Activated Mo2c. *J. Catal.* **1993,** *143*, 249−261.

(50) Blekkan, E. A.; Cuong, P. H.; Ledoux, M. J.; Guille, J. Isomerization of N-Heptane on an Oxygen-Modified Molybdenum Carbide Catalyst. *Ind. Eng. Chem. Res.* **1994,** *33*, 1657−1664.

(51) Delporte, P.; Meunier, F.; Phamhuu, C.; Vennegues, P.; Ledoux, M. J.; Guille, J. Physical Characterization of Molybdenum Oxycarbide Catalyst - Tem, Xrd and Xps. *Catal. Today* **1995,** *23*, 251−267.

(52) Oyama, S. T.; Delporte, P.; PhamHuu, C.; Ledoux, M. J. Tentative Structure of Molybdenum Oxycarbide. *Chem. Lett.* **1997**, 949−950.

(53) Pistonesi, C.; Juan, A.; Farkas, A. P.; Solymosi, F. Dft Study of Methanol Adsorption and Dissociation on Beta-Mo2c(001). *Surf. Sci.* **2008,** *602*, 2206−2211.

(54) Shi, Y.; Yang, Y.; Li, Y. W.; Jiao, H. J. Theoretical Study About Mo2c(101)-Catalyzed Hydrodeoxygenation of Butyric Acid to Butane for Biomass Conversion. *Catal. Sci. Technol.* **2016,** *6*, 4923−4936.

(55) Kunkel, C.; Vines, F.; Illas, F. Transition Metal Carbides as Novel Materials for Co2 Capture, Storage, and Activation. *Energ. Environ. Sci.* **2016,** *9*, 141−144.

(56) Wolden, C. A.; Pickerell, A.; Gawai, T.; Parks, S.; Hensley, J.; Way, J. D. Synthesis of Beta-Mo2c Thin Films. *ACS Appl. Mat. Interf.* **2011,** *3*, 517−521.
(57) Turner, R.; Fuierer, P. A.; Newnham, R.; Shrout, T. Materials for High Temperature Acoustic and Vibration Sensors: A Review. *Appl. Acoust.* **1994,** *41*, 299−324.
(58) Ren, J.; Huo, C. F.; Wang, J. G.; Li, Y. W.; Jiao, H. J. Surface Structure and Energetics of Oxygen and Co Adsorption on Alpha-Mo2c(0001). *Surf. Sci.* **2005,** *596*, 212−221.
(59) Shi, X. R.; Wang, J. G.; Hermann, K. Co and No Adsorption and Dissociation at the B-Mo2c(0001) Surface: A Density Functional Theory Study. *J. Phys. Chem. C* **2010,** *114*, 13630−13641.
(60) Posada-Perez, S.; Vines, F.; Ramirez, P. J.; Vidal, A. B.; Rodriguez, J. A.; Illas, F. The Bending Machine: Co2 Activation and Hydrogenation on Delta-Moc(001) and Beta-Mo2c(001) Surfaces. *Phys. Chem. Chem. Phys.* **2014,** *16*, 14912−14921.
(61) Politi, J. R. D.; Vines, F.; Rodriguez, J. A.; Illas, F. Atomic and Electronic Structure of Molybdenum Carbide Phases: Bulk and Low Miller-Index Surfaces. *Phys. Chem. Chem. Phys.* **2013,** *15*, 12617−12625.
(62) Kresse, G.; Furthmuller, J. Efficient Iterative Schemes for Ab Initio Total-Energy Calculations Using a Plane-Wave Basis Set. *Phys. Rev. B* **1996,** *54*, 11169−11186.
(63) Perdew, J. P.; Burke, K.; Ernzerhof, M. Generalized Gradient Approximation Made Simple. *Phys. Rev. Lett.* **1996,** *77*, 3865−3868.
(64) Kresse, G.; Joubert, D. From Ultrasoft Pseudopotentials to the Projector Augmented-Wave Method. *Phys. Rev. B* **1999,** *59*, 1758−1775.
(65) Epicier, T.; Dubois, J.; Esnouf, C.; Fantozzi, G. Neutron Powder Diffraction Studies of Transition Metal Hemicarbides M2c(1-X) - Ii. In Situ High Temperature Study on W2c(1-X_) and Mo2c(1-X). *Acta Met.* **1988,** *38*, 1903−1921.
(66) Stevanovic, V. Sampling Polymorphs of Ionic Solids Using Random Superlattices. *Phys. Rev. Lett.* **2016,** *116*, 075503.
(67) *https://github.com/pylada/pylada-light*.
(68) Chuang, F. C.; Ciobanu, C. V.; Shenoy, V. B.; Wang, C. Z.; Ho, K. M. Finding the Reconstructions of Semiconductor Surfaces Via a Genetic Algorithm. *Surf. Sci.* **2004,** *573*, L375−L381.
(69) Ciobanu, C. V.; Predescu, C. Reconstruction of Silicon Surfaces: A Stochastic Optimization Problem. *Phys. Rev. B* **2004,** *70*, 085321.
(70) Trimarchi, G.; Freeman, A. J.; Zunger, A. Predicting Stable Stoichiometries of Compounds Via Evolutionary Global Space-Group Optimization. *Phys. Rev. B* **2009,** *80*, 092101.
(71) Glass, C. W.; Oganov, A. R.; Hansen, N. Uspex - Evolutionary Crystal Structure Prediction. *Comput. Phys. Commun.* **2006,** *175*, 713−720.
(72) Ciobanu, C. V.; Shenoy, V. B.; Wang, C. Z.; Ho, K. M. Structure and Stability of the Si(105) Surface. *Surf. Sci.* **2003,** *544*, L715−L721.
(73) Han, J. W.; Li, L. W.; Sholl, D. S. Density Functional Theory Study of H and Co Adsorption on Alkali-Promoted Mo2c Surfaces. *J. Phys. Chem. C* **2011,** *115*, 6870−6876.
(74) Wang, T.; Liu, X. W.; Wang, S. G.; Huo, C. F.; Li, Y. W.; Wang, J. G.; Jiao, H. J. Stability of Beta-Mo2c Facets from Ab Initio Atomistic Thermodynamics. *J. Phys. Chem. C* **2011,** *115*, 223560−22368.
(75) Bartholomew, C. H.; Farrauto, R. J. *Fundamentals of Industrial Catalytic Processes*; Second ed.; Wiley-Interscience: Hoboken, New Jersey, 2006.
(76) Edamoto, K.; Sugihara, M.; Ozawa, K.; Otani, S. Oxidation Process of Mo2c(0001) Studied by Photoelectron Spectroscopy. *Appl. Surf. Sci.* **2004,** *237*, 498−502.
(77) *https://materialsproject.org/*.
(78) Izumi, F.; Komma, K. Three-Dimensional Visualization in Powder Diffraction. *Solid State Phenom.* **2007,** *130*, 15−20.
(79) Bader, R. F. W. *Atoms in Molecules: A Quantum Theory*; Oxford University Press: London, 1994.

(80) Sanville, E.; Kenny, S. D.; Smith, R.; Henkelman, G. Improved Grid-Based Algorithm for Bader Charge Allocation. *J. Comput. Chem.* **2007,** *28*, 899−908.
(81) Frank, B.; Cotter, T. P.; Schuster, M. E.; Schogl, R.; Trunschke, A. Carbon Dynamics on the Molybdenum Carbide Surface During Catalytic Propane Dehydrogenation. *Chem. Eur. J.* **2013,** *19*, 16938−16945.
(82) Schaidle, J. A.; Lausche, A. C.; Thompson, L. T. Effects of Sulfur on Mo2c and Pt/Mo2c Catalysts: Water Gas Shift Reaction. *J. Catal.* **2010,** *272*, 235−245.
(83) Lu, Q.; Chen, C.-J.; Luc, W.; Chen, J. G.; Bhan, A.; Jiao, F. Ordered Mesoporous Metal Carbides with Enhanced Anisole Hydrodeoxygenation Selectivity. *ACS Catal.* **2016,** *6*, 3506−3514.
(84) Solymosi, F.; Oszko, A.; Bansagi, T.; Tolmacsov, P. Adsorption and Reaction of Co2 on Mo2c Catalyst. **2002,** *106*, 9613.

**SUPPORTING INFORMATION FOR PUBLICATION:**

# Thermodynamic Stability of Molybdenum Oxycarbides Formed from Orthorhombic $Mo_2C$ in Oxygen-rich Environments

S.R.J. Likith,[a,†] C. A. Farberow,[b,*] S. Manna,[a,†] A. Abdulslam,[a] V. Stevanović,[c,d] D. A. Ruddy,[b] J. A. Schaidle,[b] D. J. Robichaud,[b,*] and C.V. Ciobanu[a,*]

[a]*Department of Mechanical Engineering, Colorado School of Mines, Golden, CO 80401, USA*
[b]*National Bioenergy Center, National Renewable Energy Laboratory, Golden, CO 80401, USA*
[c]*National Renewable Energy Laboratory, Golden, CO 80401, USA*
[d]*Department of Metallurgical and Materials Engineering, Colorado School of Mines, Golden, CO, 80401, USA*

## 1. Top views for the surface configurations in the phase diagram

Figures S1 through S4 present the low energy configurations that appear in Fig. 2 of the main text. As explained in the text, these may not necessarily be the lowest possible configurations, but are configurations that are representative in terms of the content of C and O in above and below the top Mo layer. For the sake of clarity, only top configurations are shown. The rectangle represents the surface unit cell (5.25067 Å × 6.07256 Å).

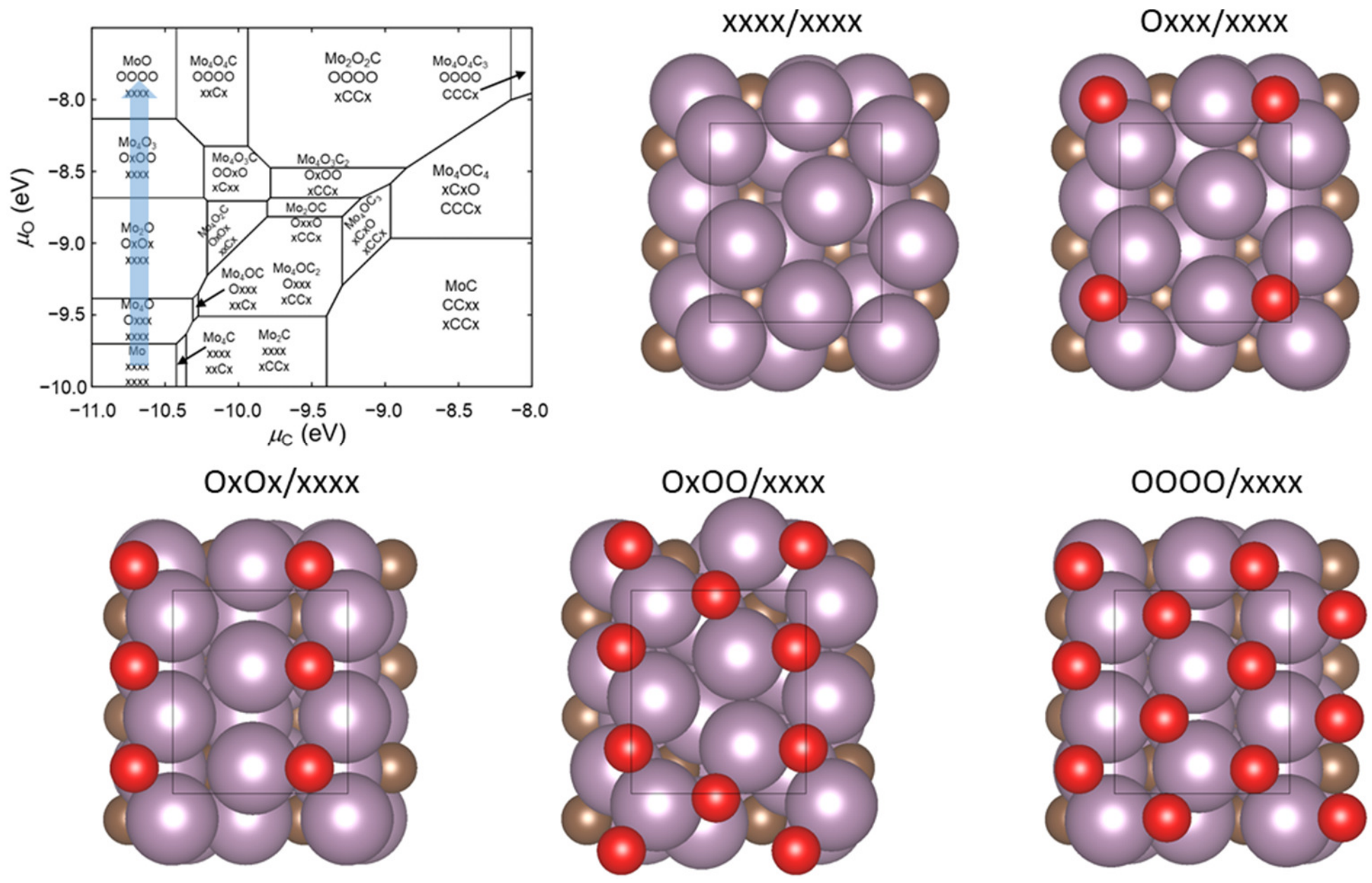

**Figure S1.** Surface configurations for low chemical potential of carbon (no C atoms).

† These authors made equal contributions to the results in this work
* To whom correspondence may be addressed: carrie.farberow@nrel.gov (C. A. Farberow), david.robichaud@nrel.gov (D. J. Robichaud), and cciobanu@mines.edu (C.V. Ciobanu)

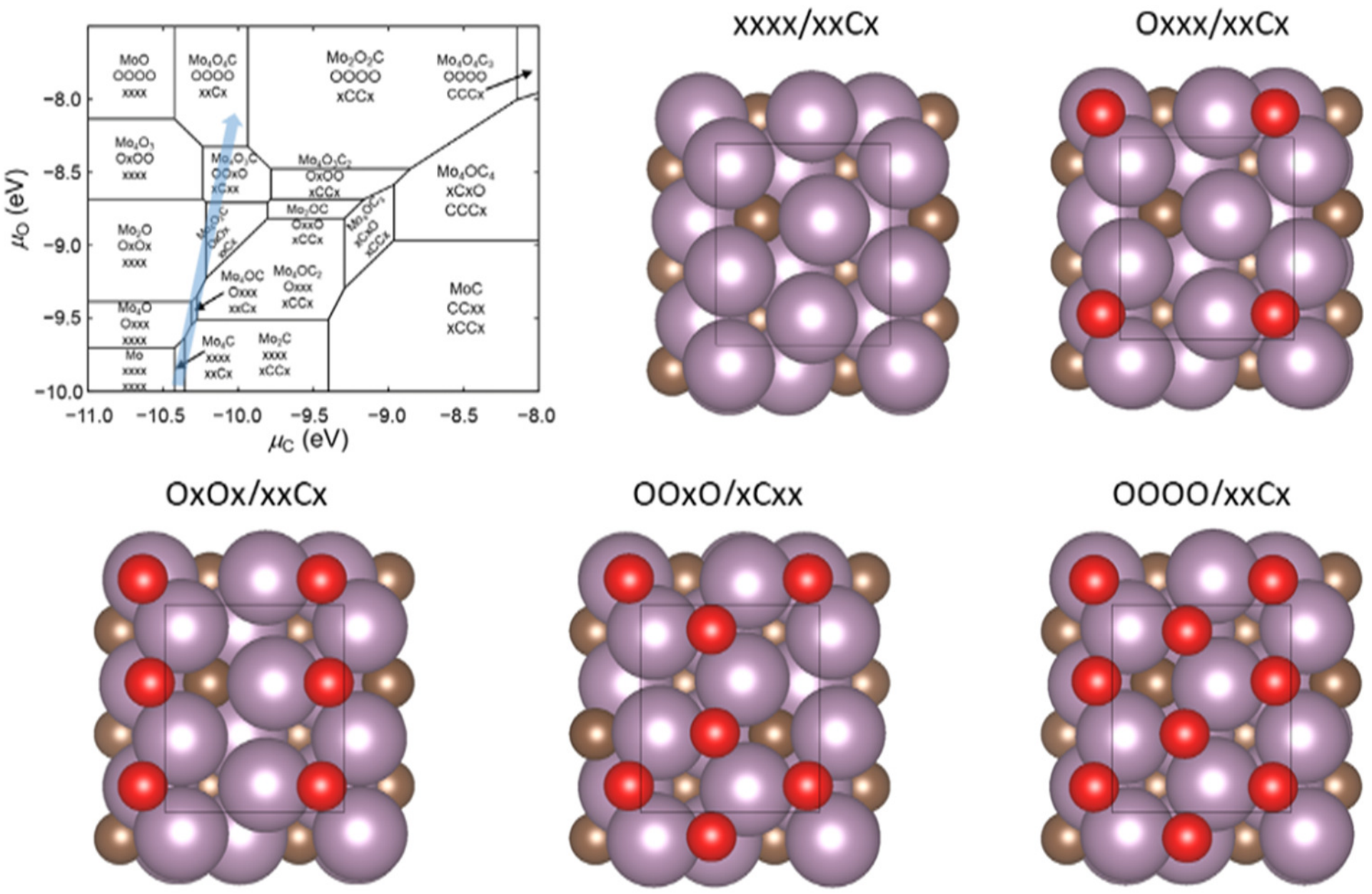

**Figure S2.** Surface configurations for intermediate chemical potential of carbon (one C atom per unit cell below the Mo layer).

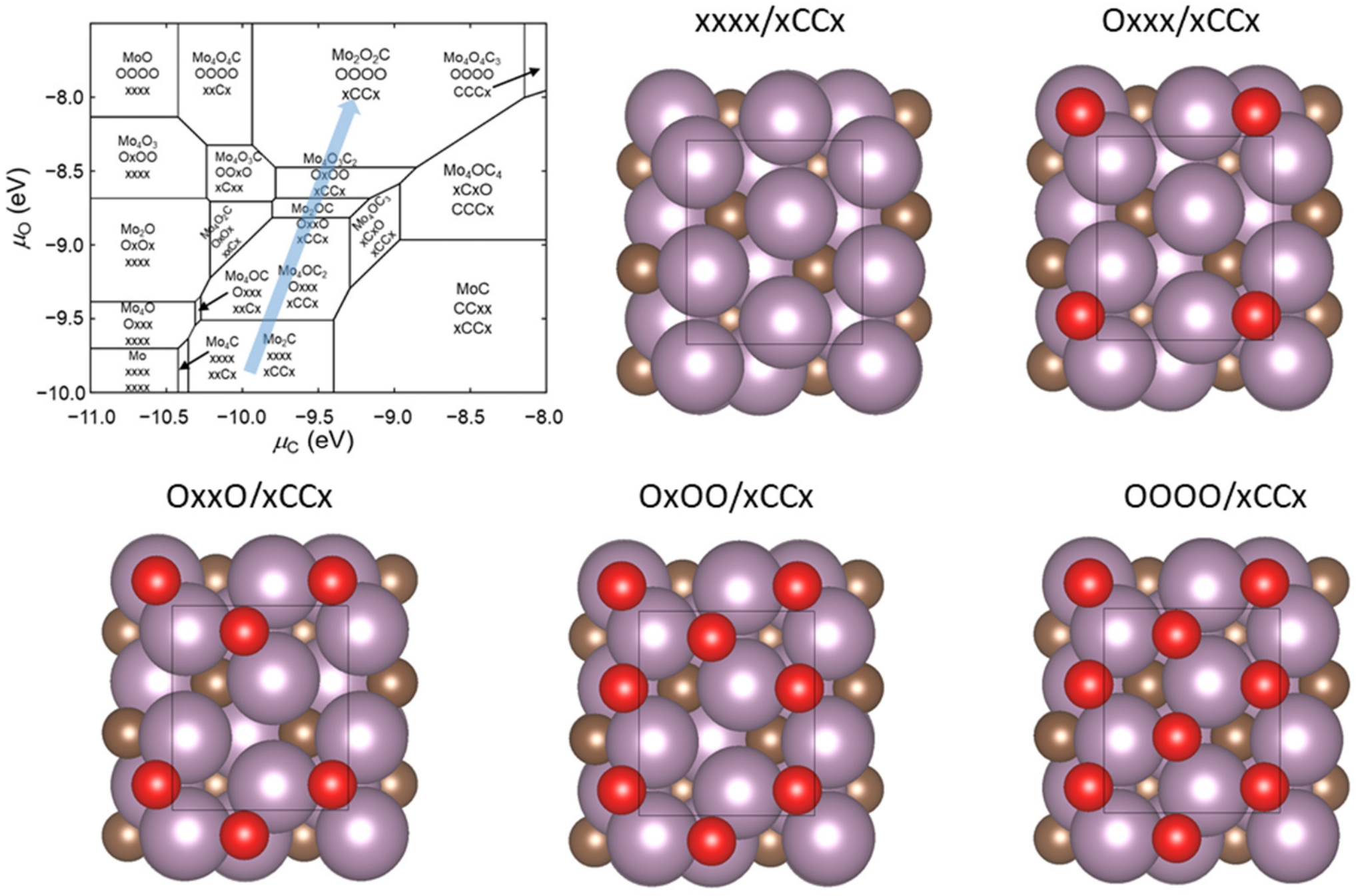

**Figure S3.** Surface configurations for intermediate chemical potential of carbon (two C atoms per unit cell below the Mo layer).

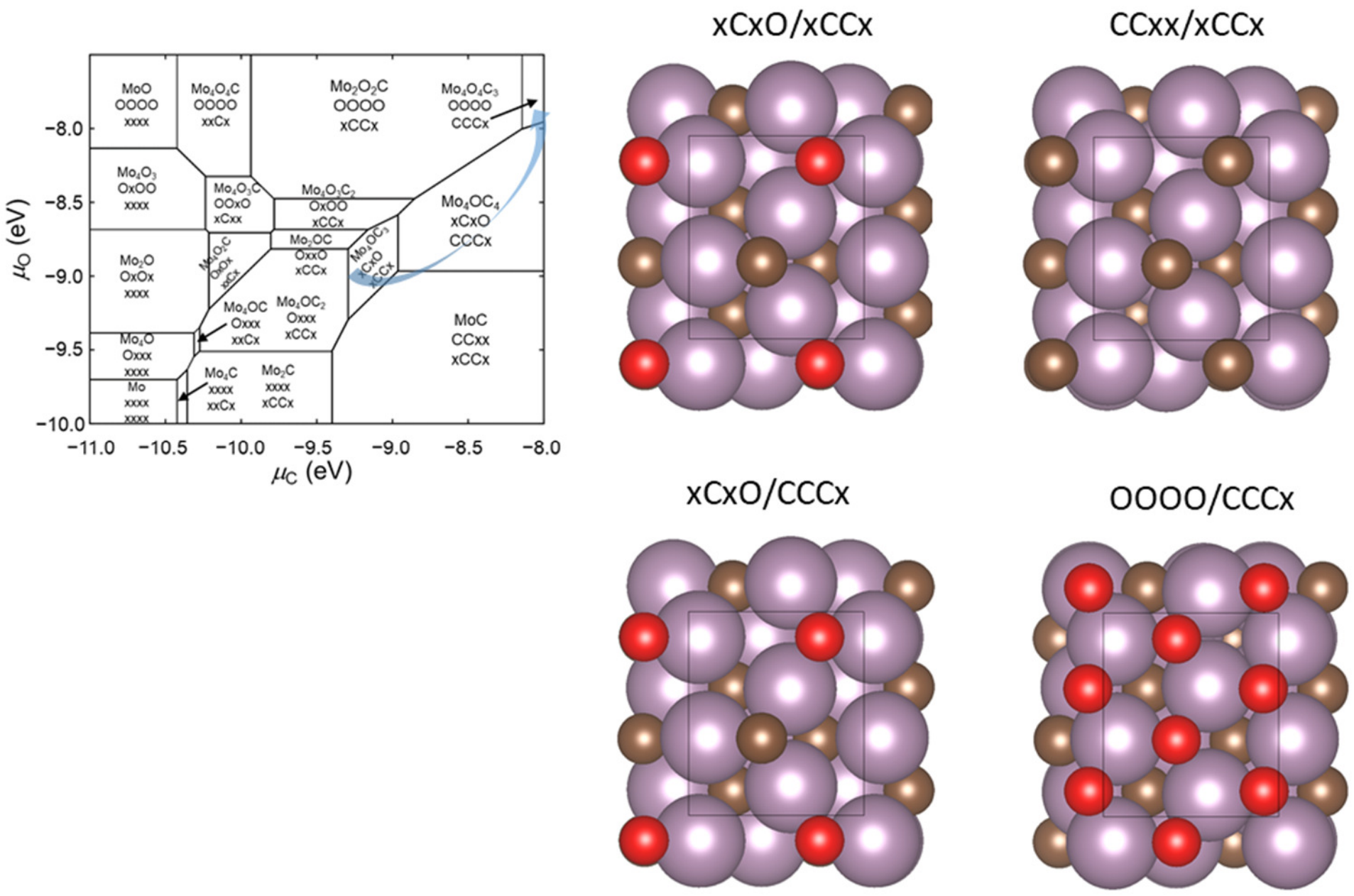

**Figure S4.** Surface configurations for high chemical potential of carbon (two or three C atoms per unit cell below the Mo layer).

## 2. Chemical reactions leading to the deposition of carbon and oxygen

In catalytic reforming, the steam methane reforming reaction (SMR) is usually followed by the water gas shift reaction (WGS):

$CH_4+H_2O \leftrightarrow CO+3H_2$ (SMR)

$CO+H_2O \leftrightarrow H_2+CO_2$ (WGS)

Along with water, the products of these reactions lead to diverse means for the deposition of C and O on the catalyst. Carbon can be deposited (dep) via the following reactions:

$CO+H_2 \leftrightarrow H_2O + C$ (dep) (syngas deposition, SD)

$2CO \leftrightarrow CO_2 + C$ (dep) (CO disproportionation, COD)

Oxygen can be deposited on, or incorporated in, the catalyst through the decomposition of water or carbon dioxide, and directly through oxygen dissociation:

$H_2O \leftrightarrow H_2 + O$ (dep) (water decomposition, WD)

$CO_2 \leftrightarrow CO + O$ (dep) (carbon dioxide decomposition, CDD)

These source reactions for carbon (SD, COD) and oxygen (WD, CDD) at thermodynamic equilibrium, serve for computing the *chemical potential of the deposited species (C or O) solely in terms of gas phase potentials*:

$$\mu_{\mathrm{C}} = \mu_{\mathrm{CO}} + \mu_{\mathrm{H_2}} - \mu_{\mathrm{H_2O}} \qquad \text{(A.1, for SD as the source of C)}$$

$$\mu_{\mathrm{C}} = 2\mu_{\mathrm{CO}} - \mu_{\mathrm{CO_2}} \qquad \text{(A.2, for COD as the source of C)}$$

$$\mu_{\mathrm{O}} = \mu_{\mathrm{H_2O}} - \mu_{H_2} \qquad \text{(A.3, for WD as the source of O)}$$

$$\mu_{\mathrm{O}} = \mu_{\mathrm{CO_2}} - \mu_{\mathrm{CO}} \qquad \text{(A.4, for CDD as the source of O)}$$

For a reactant or product in the gas phase [i.e., situated on the right hand side of Eqs (A.1)-(A.4)] the chemical potential can be determined from the experimental enthalpy $H^{\mathrm{o}}$ and entropy $S^{\mathrm{o}}$ in thermodynamic tables, to which we supplement the energy at zero Kelvin (specifically, the bond energy plus the zero-point-energy of vibrations). The tables used were the NIST –JANAF Thermochemical Tables, http://kinetics.nist.gov/janaf/ . For each of the gaseous species on the right-hand side of Eqs. (A.1)-(A.4), the chemical potential at temperature *T* and pressure *p* given by[1]

$$\mu\,(T,p) = \mu^{\mathrm{o}}(T) + k_{\mathrm{B}}T\ln\frac{p}{p_0} = \qquad \text{(A.5)}$$

$$= E_0 + \left(H^{\mathrm{o}}(T) - H^{\mathrm{o}}(0\mathrm{K})\right) - TS^{\mathrm{o}}(T) + k_{\mathrm{B}}T\ln\frac{p}{p_0}\ , \qquad \text{(A.6)}$$

where the superscript o refers to the standard pressure of $p_0$ = 1 atm, $E_0$ is the sum between the bond energy and zero-point energy, and $k_{\mathrm{B}}$ is the Boltzmann constant. For determining $E_0$ (Table S1), we used the Amsterdam Density Functional (ADF) software package (ADF),[2] with the B3LYP exchange-energy functional,[3-4] no frozen cores, and augmented triple-zeta polarization functions (AUG/ATZP). The main reason for using B3LYP is that it performs better than PBE for molecules,[5] while for condensed systems (bulk, surfaces) PBE is more appropriate. Since in Eqn. (A.6) we are combining theoretical ($E_0$) and experimental (rest of the terms in (A.6)) quantities to determine chemical potentials at finite temperatures, it is important to use an accurate functional for molecular systems.

**Table S1**. B3LYP bond energy and zero-point energy for the gaseous species in Eqs. (A.1-A.4)

| Gas species | Bond Energy (eV) | Zero-Point Energy (eV) | $E_0$ (eV) |
|---|---|---|---|
| CO | -18.0053 | 0.1361 | -17.8691 |
| $H_2$ | -7.7149 | 0.2728 | -7.4421 |
| $H_2O$ | -17.0270 | 0.5766 | -16.4503 |
| $CO_2$ | -27.7118 | 0.3144 | -27.3974 |

The reaction conditions[6] used in calculating the chemical potentials for the deposited C and O atoms are listed in Table S2. The partial pressures in the last two columns of Table S2 are determined from typical inlet out outlet gas compositions[6] for SMR and WGS.

**Table S2.** Reaction conditions considered in the chemical potential calculations

| Reaction | Temperature range, in/out (°C) | Pressure range, in/out (bar) | Partial pressures for source reactions, in (bar) | Partial pressures for source reactions, out (bar) |
|---|---|---|---|---|
| SD / SMR | 700 – 1000 | 20 – 40 | 0.002 – 0.004 (CO)<br>0.002 – 0.004 ($H_2$)<br>13.4 – 26.8 ($H_2O$) | 2.83 – 5.66 (CO)<br>13.96 – 27.92 ($H_2$)<br>1.13 – 2.26 ($H_2O$) |
| COD / SMR | 700 – 1000 | 20 – 40 | 0.002 – 0.004 (CO)<br>0.002 – 0.004 ($CO_2$) | 2.83 – 5.66 (CO)<br>1.51 – 3.02 ($CO_2$) |
| WD / SMR | 700 – 1000 | 20 – 40 | 0.002 – 0.004 ($H_2$)<br>13.4 – 26.8 ($H_2O$) | 13.96 – 27.92 ($H_2$)<br>1.13 – 2.26 ($H_2O$) |
| CDD / SMR | 700 – 1000 | 20 – 40 | 0.002 – 0.004(CO)<br>0.002 – 0.004 ($CO_2$) | 2.83 – 5.66 (CO)<br>1.51 – 3.02 ($CO_2$) |
| SD / WGS | 200 – 450 | 1 – 30 | 0.0107 – 3.21 (CO)<br>0.5 – 1.5 ($H_2$)<br>0.286 – 8.57 ($H_2O$) | 0.0008 – 0.024 (CO)<br>0.62 – 18.56 ($H_2$)<br>0.214 – 6.42 ($H_2O$) |
| COD / WGS | 200 – 450 | 1 – 30 | 0.0107 – 3.21 (CO)<br>0.1 – 3.0 ($CO_2$) | 0.0008 – 0.024 (CO)<br>0.166 – 4.972 ($CO_2$) |
| WD / WGS | 200 – 450 | 1 – 30 | 0.5 – 1.5 ($H_2$)<br>0.286 – 8.57 ($H_2O$) | 0.62 – 18.56 ($H_2$)<br>0.214 – 6.42 ($H_2O$) |
| CDD / WGS | 200 – 450 | 1 – 30 | 0.0107 – 3.21 (CO)<br>0.1 – 3.0 ($CO_2$) | 0.0008 – 0.024 (CO)<br>0.166 – 4.972 ($CO_2$) |

We have calculated [Eqn. (A.6)] the chemical potential ranges for deposited carbon and oxygen for the specific conditions listed above, and tabulated them in Table S3.

**Table S3.** Chemical potential ranges calculated for conditions associated with SMR and WGS.

| | SMR | WGS |
|---|---|---|
| SD, $\mu_C$ (eV) | [-12.78, -10.38] | [-10.33, -9.78] |
| COD, $\mu_C$ (eV) | [-11.49, -9.98] | [-10.44, -9.11] |
| WD, $\mu_O$(eV) | [-10.01, -7.67] | [-9.49, -8.70] |
| CDD, $\mu_O$(eV) | [-9.94, -9.75] | [-9.67, -9.34] |

**3. Oxidation states from scaled Bader charges**

The increase in the average coordination of Mo atoms with oxygen should result in changes in the oxidation states of the Mo. The oxidation states of Mo atoms are not directly computable quantities, but we can assess them from scaling the net charge on every Mo, i.e. difference between the ionic core charge and the Bader electrons.[7-8] The Bader charges are converted into oxidation states by scaling the charge on each ion such that the average charge across all oxygens is $-2$. In the case of the pure molybdenum carbide, we use $-4$ as the reference valence for carbon. This method predicts correct oxidation states for Mo in $Mo_2C$, $MoO_2$, and $MoO_3$. Figure S5 shows the computed Mo oxidation states for all atoms of the oxycarbides described in the main text. The oxidation states of Mo in the orthorhombic $Mo_2C$ and in the oxides ($MoO_2$ and $MoO_3$) are also reported for comparison. There is an (artificial) drop in oxidation states from $Mo_2C$ ($y/x = 0$) to oxycarbides with low oxygen content ($0 < y/x \leq 1$), which is due to changing the reference states from C ($-4$) to O ($-2$). The valence of Mo in oxycarbides disperses over an interval, rather than assuming fixed values. For compositions with $y/x < 1$, the Mo oxidation states are between 1 and 2.7, whereas for compositions with $y/x > 1$ the Mo oxidation states are between 2 and 4. Most of these states are distinct from the oxidation state of Mo in the parent $Mo_2C$ phase (+2) and should appear in experimental XPS spectra.

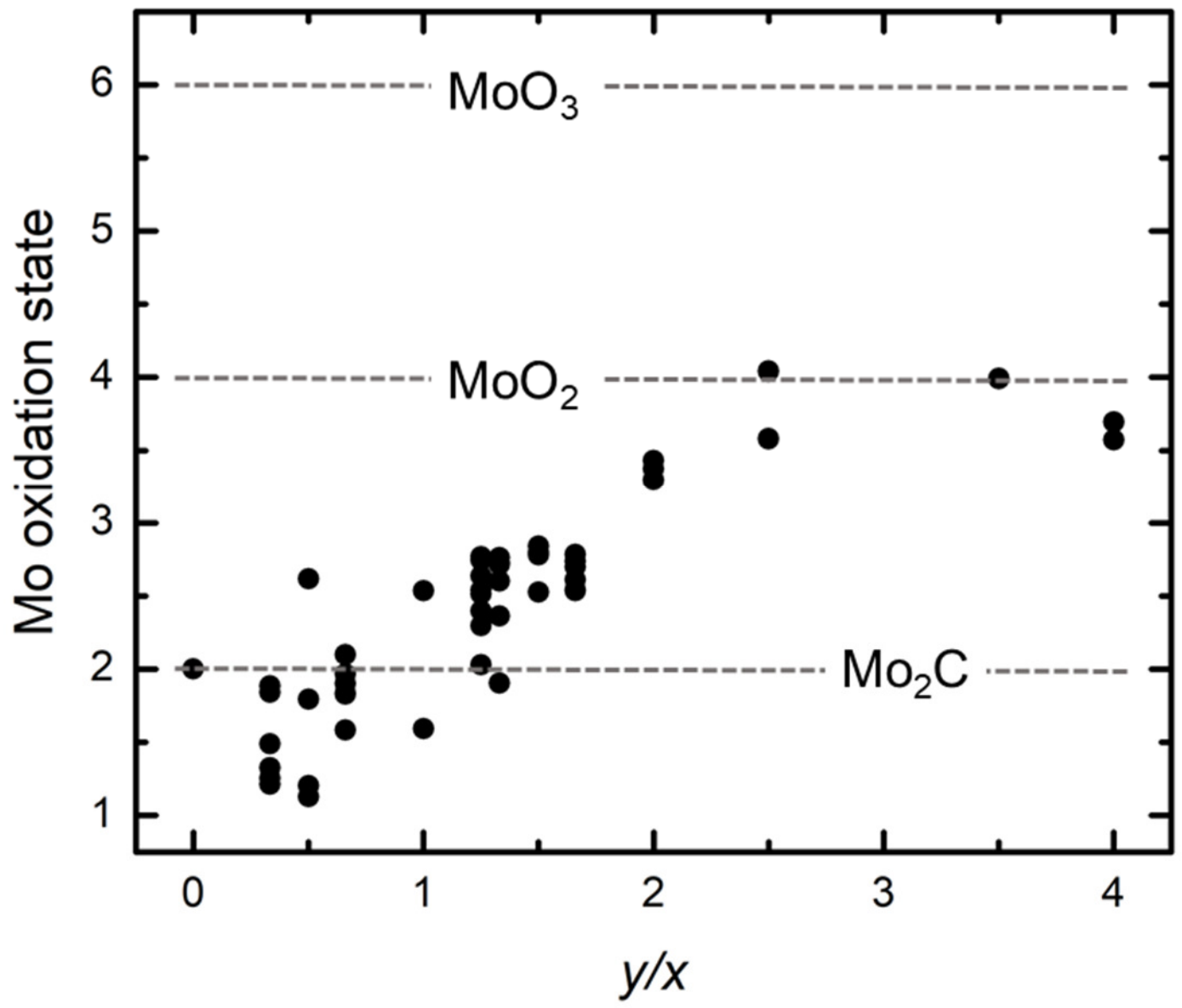

**Figure S5.** Oxidation states of Mo atoms for each of the low-energy oxycarbide phases obtained from the structural search. For comparison, the dashed lines show the standard states of Mo atoms in the pure carbide and the two crystalline oxides of Mo.

## References

(1) McQuarrie, D. A. *Statistical Mechanics*; University Science Books: London, 2000.

(2) ADF2013; SCM; http://www.scm.com (Accessed July 2017). Theoretical Chemistry, Vrije Universiteit, Amsterdam, the Netherlands,.

(3) Becke, A. D. Density Functional Thermochemistry. 3. The Role of Exact Exchange *J. Chem. Phys.* **1993,** 98, 5648−5652.

(4) Lee, C.; Yang, W.; Parr, R. G. Development of the Colle-Salvetti Correlation-Energy Formula into a Functional of the Electron Density. *Phys. Rev. B* **1988,** 37, 785−789.

(5) Staroverov, V. N.; Scuseria, G. E.; Tao, J. M.; Perdew, J. P. Comparative Assessment of a New Nonempirical Density Functional: Molecules and Hydrogen-Bonded Complexes. *J. Chem. Phys.* **2003,** 119, 12129−12137.

(6) Bartholomew, C. H.; Farrauto, R. J. *Fundamentals of Industrial Catalytic Processes*; Second ed.; Wiley-Interscience: Hoboken, New Jersey, 2006.

(7) Bader, R. F. W. *Atoms in Molecules: A Quantum Theory*; Oxford University Press: London, 1994.

(8) Sanville, E.; Kenny, S. D.; Smith, R.; Henkelman, G. Improved Grid-Based Algorithm for Bader Charge Allocation. *J. Comput. Chem.* **2007,** 28, 899−908.